\documentclass[Afour,sageh,times]{sagej}

\usepackage{moreverb,url}

\usepackage[colorlinks,bookmarksopen,bookmarksnumbered,citecolor=red,urlcolor=red]{hyperref}

\newcommand\BibTeX{{\rmfamily B\kern-.05em \textsc{i\kern-.025em b}\kern-.08em
T\kern-.1667em\lower.7ex\hbox{E}\kern-.125emX}}

\def\volumeyear{2016}

\begin{document}

\runninghead{Nespeca et al.}
\title{Learning to connect in action: Measuring and understanding the emergence of boundary spanners in volatile times}

\author{Vittorio Nespeca\affilnum{1} and Tina Comes \affilnum{1} and Frances Brazier\affilnum{1}}

\affiliation{\affilnum{1}Delft University of Technology, NL}

\corrauth{Vittorio Nespeca, Delft University of Technology,
Delft, South Holland, 2628 BX, The Netherlands.}

\email{v.nespeca@tudelft.nl}

\begin{abstract}
        Collective intelligence of diverse groups is key for tackling many of today's grand challenges such as fostering resilience and climate change adaptation. Given increasing societal fragmentation, information exchange across such diverse groups is crucial for collective intelligence, especially in volatile environments. To facilitate inter-group information exchange, Informational Boundary Spanners (IBSs) as pivotal information exchange 'hubs' are promising. However, the mechanisms that drive the emergence of IBSs remain poorly understood. To address this gap there is first a need for a method to identify and measure the emergence of IBSs. Second, an Agent-Based Modelling (ABM) framework is not available to systematically study mechanisms for the emergence of IBSs in volatile environments through the method. Third, even though the ability to learn who provides high-quality information is thought to be essential to explain the emergence of IBSs, a rigorous test of this mechanism is missing. The learning mechanism is formalized using an ABM framework, with the model's outputs analyzed through the proposed method for measuring the emergence of IBSs. To illustrate both the measurement method and the learning mechanism, we present a case study focused on information sharing in the volatile environment of a disaster. The study shows that learning constitutes a mechanism for the emergence of effective IBSs in (a) low-volatility environments characterised by low uncertainty and (b) in high-volatility environments characterised by rapid change \textit{if} the number of inter-group connections is sufficient. With the method and model, this paper aims to lay the foundations for exploring mechanisms for the emergence of IBSs that facilitate inter-group information exchange. This article contributes to the field of collective intelligence by providing the essential elements for measuring and understanding the emergence of IBSs and by exploring the effect of learning on their emergence in volatile environments.

\end{abstract}


\maketitle

\section*{Significance statement} 
In today's rapidly changing and uncertain world, effective communication and collaboration are crucial between different groups including governmental organizations, NGOs, and communities. This paper focuses on effective exchange of information within and between these groups. Individual actors known as Informational Boundary Spanners (IBSs) play a key role in this process. Yet, there is so far no method to quantitatively measure the emergence of individual IBSs, and as such no analytical way to systematically test the mechanisms that drive the emergence of IBSs. Measuring and understanding the conditions and underlying mechanisms that influence the emergence of IBSs, and their success, is the focus of this paper. 

This paper (a) proposes a method to quantitatively measure the number of actors that take on the role of IBSs and (b) shows how this method can be used in combination with computer simulations to study mechanisms for the emergence of IBSs. In particular, the ability of all actors to learn which actors provide information of high quality is explored as a mechanism for the emergence of IBSs under different conditions.

Results show that learning can facilitate more actors into growing into the role of IBSs that effectively convey information across groups. This process is most effective when information is consistently provided by information sources that are stable and do not change over time. Further, in environments characterized by rapid change, a high number of inter-group contacts (20 or more) is vital for the development of IBSs, to distribute the informational exchange burden and improve communication efficiency.

These findings have two main practical implications. First, to foster effective inter-group information exchange in volatile environments it is key to broadly crowd-source information and widely disseminate it through stable channels. Second, the availability of trusted connections across groups is crucial to support effective inter-group information exchange even in conditions of high volatility. This, in turn, mandates that trusted connections between groups are established and maintained over time: a true societal challenge. 

\section{Introduction}
Policies and practices for ensuring a sustainable, resilient, and climate-adaptive future rely on collective intelligence across diverse groups \citep{norris_community_2008, malone2018superminds, rockstrom2023shaping, perrone2023stakeholder}. 
Collective intelligence is defined as the shared problem-solving ability that arises from the interaction and combined efforts of a group of individuals, leading to the effective accomplishment of one or more goals  \citep{malone2018superminds, flack2022editorial}. Fostering collective intelligence strongly relies on the exchange of information among the individuals belonging to the different groups \citep{leonard2022collective, ballou2007evolution, Malone_Crowston_interdisciplinary, fiala2005information}. The need for information sharing is even more prominent in volatile (i.e., rapidly changing and uncertain) environments, where the groups (e.g., governmental organizations, NGOs, and communities) must continually adapt while maintaining coordination of their activities. Disaster response exemplifies a situation in which multiple groups operate in a volatile environment and need to exchange information to coordinate and respond effectively \citep{kapucu2005interorganizational, bharosa_challenges_2010, comes2020coordination, nespeca2020towards}. 

Actors within these groups exchange information with each other and with actors in other groups. Some actors do so more successfully than others: over time they become hubs for inter-group information exchange or Informational Boundary Spanners (IBSs) \citep{tushman1977special, ancona1992bridging,kapucu2006interagency, ballou2007evolution, zagorecki_interorganizational_2009, marrone2010team, altay_information_2013, altay2014challenges, van2014effects, bharosa_principle-based_2015}. \citet{levina2005emergence} find that the formal appointment of this role as IBS, such as through a mandate, does not suffice to ensure effective information exchange. 
Instead, the role of IBSs in facilitating information sharing among groups emerges through dynamic interactions within and between these groups.
More specifically, fostering the emergence of IBSs requires consideration of the interplay between individual group members (micro level), their groups (meso level), and networked interactions among groups (macro level) \citep{marrone2007multilevel,marrone2010team}. This emergent process is contingent on specific conditions at the different levels. For example, it is key that at the micro level an IBS develops an interest in carrying out boundary spanning. Further, the formal nomination of an actor as an IBSs at the group (meso) level can support his/her emergence as such but is not mandatory. And, finally, formal or informal recognition  of boundary spanners' authority to negotiate on behalf of their group also plays as key role in the emergence of boundary spanners when considering networked interactions among groups (macro level). 

While there are some initial empirical insights into the emergence of informational boundary spanning at the micro, meso, and macro levels and mechanisms that drive their emergence, there is thus far no quantitative formalisation and analytical modeling framework that studies the emergence of IBSs via computational experiments. This paper argues that there are three major gaps in the literature: (i) a formalised method to quantitatively analyse how IBSs effectively conveying information across groups emerge; (ii) an analytical modeling framework to analyse and understand the mechanisms behind the emergence of effective IBSs under different conditions of environmental volatility; and (iii) a deeper understanding of the effects of learning on the emergence of IBSs in volatile environments.

Case study research has been invaluable to study boundary spanning  \citep{ancona1992bridging, levina2005emergence, marrone2010team, lifshitz2018dismantling}. Yet, Agent-Based Modeling (ABM) can complement and enhance case study research by facilitating the systematic analysis and comparison of results from several case studies by capturing interactions from micro to macro levels within a unified modeling framework. This is particularly relevant when investigating emergent mechanisms resulting from complex interactions across multiple levels \citep{adam_modelling_2017, antosz2022sensemaking}, as seen with IBSs \citep{marrone2010team}. For instance, by incorporating insights from case studies on individual behavior at the micro level into an ABM, researchers can simulate a system's macro-level behavior, assessing whether the simulated emergent patterns replicate and explain empirical observations from other studies \citep{adam_modelling_2017}. This approach enables the replication, testing of consistency, and extension of case study research findings and provides directions for further case studies \citep{tubaro2010ethnographic, castellani2019case, nespeca_methodology_2023}. 

A fundamental step in enabling a combination of case study research and ABM when studying the emergence of IBSs is to design a quantitative method for measuring boundary spanning on an individual or micro level. This involves \textit{identifying individuals who emerge as IBSs} by effectively providing the information needed to the groups who need it. Identifying emergent IBSs is vital for examining the interplay between the micro, meso, and macro level conditions that promote such emergence \citep{levina2005emergence}.  While  several quantitative methods for IBSs have been provided for measuring informational boundary spanning \citep{ancona1992bridging, hansen1999search, cummings2004work, marrone2010team, van_osch_towards_2016}, these methods typically focus on boundary spanning at the level of one or multiple groups rather than at the level of the individual actor. Specifically, some methods focus on assessing the volume of boundary spanning (e.g., through the frequency of communication) vis a vis operational performance (e.g., time required for project completion) at the group level \citep{hansen1999search,cummings2004work, zagorecki_interorganizational_2009}. Other methods, directly measure the success of IBSs in retrieving information for their group in a timely manner \citep{ancona1992bridging, marrone2007multilevel, van_osch_towards_2016}. Finally, \citet{zagorecki_interorganizational_2009} monitors network characteristics (e.g., the average distance between nodes) and the volume of inter-group information exchanged to study the level of informational boundary spanning occurring among groups. However, none of these approaches measures how many and which individual agents emerge as effective IBSs. As such, a method to measure the emergence of effective IBSs at micro-level is missing, leading to a lack of understanding in the mechanisms behind the emergence of IBSs. 

Such a method provides a basis to build an ABM framework enabling to simulate inter-group information exchange and capture the simulation outputs required to systematically analyse, test, and understand at micro-level mechanisms for the emergence of IBSs under varying circumstances. The characteristics of the external environment, in which the groups operate, are likely to play a crucial role in shaping the emergence of IBSs \citep{aldrich1977boundary, hazy_impact_2003}. The environment consists of factors external to a group's boundary that affect the decisions of individuals within the group and thereby affects the group's ability to achieve its goals \citep{lawrence1967organization, duncan1972characteristics}. Especially in volatile environments information sharing has been shown to be essential to achieve collective intelligence and coordination \citep{comes2020coordination}. Volatility is defined as the level of turbulence and uncertainty that characterize changes in the environment, where turbulence indicates the frequency of change \citep{hazy_impact_2003}, and uncertainty denotes the unpredictability in the occurrence of change \citep{duncan1972characteristics}. The emergence of boundary spanning (or lack thereof) has been primarily studied for non-volatile environments \citep{ancona1992bridging,levina2005emergence, levina2006turning, lifshitz2018dismantling}. Even though volatile situations such as social unrest, crises, or conflicts are increasingly common, little research is available on the conditions that foster the emergence of IBSs in volatile environments. \citet{hazy_impact_2003} propose an ABM framework and use it to study the effectiveness of different numbers of IBSs for varying levels of environmental volatility. However, their modeling framework and study assume a predefined number of IBSs and does not account for how and why they emerge. Further, \citet{zagorecki_interorganizational_2009} propose an ABM to study the emergence of inter-group information exchange in volatile environments. Yet, their model does not focus on capturing the emergence of IBSs at the micro level. As such, an ABM framework to study the emergence of IBSs at the micro level in volatile environments in missing. 

Further, several empirical studies have shown that actors in volatile conditions actively \textit{learn} who among their contacts provides high-quality information, subsequently adjusting their information collection preferences to align with these sources \citep{sutton2008backchannels, nespeca2020towards}. Preliminary evidence shows that this learning behavior can lead to the emergence of IBSs in volatile conditions \citep{nespeca2020towards}. However, the concrete mechanisms between learning and the emergence of IBSs remain poorly understood.

In sum, there are several gaps in the understanding of IBSs. First, a method is needed for measuring the emergence of IBSs at the micro level as key to analyse the underlying mechanisms that drive their emergence. Second, an ABM framework is missing to systematically study mechanisms for the emergence of IBSs at the micro level in volatile environments. Third, a better understanding is required of the effects of learning and the volatility of the situation on the emergence of IBSs. 

To address these gaps, this paper develops a method to measure the emergence of IBSs at the micro-level and introduces an ABM framework enabling to study the emergence of IBSs in volatile conditions through the method. Then, the method and ABM framework are used to analyse and understand learning as a mechanism that potentially results in the emergence of IBSs. To this end, two mechanisms are introduced and compared in the ABM: in the first mechanism, agents exchange information randomly. In the second mechanism, agents continually learn which sources provide the most relevant information and adjust their information collection preferences accordingly. Both mechanisms are studied for different levels of volatility and numbers of connections enabling information exchange among the considered groups during the response to a disaster. 

The remainder of this paper is structured as follows. Section 2 introduces our case study of disaster response in Jakarta as an example of a situation that requires collective intelligence. Section 3 introduces the method proposed for measuring the emergence of IBSs based on existing literature. Section 4 formulates three propositions concerning learning mechanisms for the emergence of IBSs and their impact on inter-group information exchange in volatile environments. Section 5 describes the development of an ABM that captures inter-group information exchange and provides the output required to study the emergence of IBSs through the proposed method. Section 6 illustrates the model parametrization and experimental design aimed at testing the method for capturing the emergence of IBSs and study the propositions through the ABM. Section 7 presents the results of the experiments. Section 8 discusses the implications of these results, leading to considerations regarding the correctness of the method proposed for measuring emergent IBSs, the extent to which the results support the formulated propositions, and what the findings imply for information management and collective intelligence in volatile environments. This section also presents directions for future research. Section 8 concludes the paper.




\section{Case study: Disaster response in Jakarta}
In disaster response, typically multiple and loosely connected organizations (e.g. governmental organizations and NGOs) and groups (e.g. communities) collectively operate in a highly volatile environment. These groups need to exchange information to respond effectively \citep{kapucu2005interorganizational, altay2006or, bharosa_challenges_2010, comes2020coordination, nespeca2020towards}. Information exchange in these conditions is particularly challenging as, given the sheer volume and frequency of new information produced during a disaster, the actors are likely to become overloaded with information, which impairs their ability to share and retrieve relevant information \citep{van2016improving, comes2016cognitive}. Further, disaster response is typically characterized by high uncertainty, meaning it is often difficult to predict when and from which sources relevant information will become available. As such, disaster response presents an ideal case study to understand the emergence of IBSs in volatile environments. 

This research focuses on the case study of Jakarta, Indonesia. Situated on the northwest coast of Java, the world's most populous island, Jakarta is subject to frequent flooding, primarily attributed to its rapid subsidence and ongoing urbanization processes \citep{abidin2011land, abidin2015correlation}. Jakarta also hosts diverse stakeholders including governmental organizations, NGOs, and community initiatives that need to coordinate, collaborate, and exchange information effectively. Additionally, Jakarta was chosen because, during the data collection period in 2018, numerous international organizations were present in response to the humanitarian response to the Sulawesi Earthquake. The details on the case study including data collection, analysis, and discussion are available in \citep{nespeca2020towards}.

\section{Measuring the emergence of informational boundary spanners}\label{sec:measuring-ibs} 
While several functions can be attributed to boundary spanning including information processing, external representation, negotiation, and brokering \citep{aldrich1977boundary, ancona1992bridging, fleming2007brokerage}, this paper focuses on information processing. Information processing or informational boundary-spanning is defined as the activity of searching for information that lies outside the boundary of a group (i.e. that originates in the group's external environment) to find, process, and \textit{share} new and relevant information that can enhance the knowledge of the group \citep{bharosa_netcentric_2011, lindgren2008multi, ancona1992bridging, van_osch_towards_2016}. Boundaries are the delimitation of a group or organization from its environment \citep{scott1992organizations}.  Often external information needed by a group is available from other groups. As such, performing the informational boundary spanning function entails fostering the exchange of information among groups. 

IBS candidates (actors that can become IBSs), are those who have connections across group boundaries, or 'inter-group ties'. As such, they can potentially search, find, process, and share information across groups, thus performing the informational boundary spanning function.

One approach to measuring the emergence of IBSs may be to simply count the number of IBS candidates that carry out the informational boundary-spanning function. However, while this function may be performed occasionally by many or all potential IBSs, only a few of these actors consistently perform this function and thus contribute significantly to fostering the exchange of new and relevant information across groups \citep{levina2005emergence}. As such, simply counting the number of IBS candidates that perform the informational boundary-spanning function occasionally (e.g. once or a few times) is expected to overestimate the number of IBSs that emerge. Then, measuring the emergence of IBSs requires identifying those actors that not only carry out the informational boundary spanning function but also do so with sufficient consistency to significantly enhance the exchange of new and relevant external information among groups.

\section{Understanding the emergence of informational boundary spanners} \label{sec:understanding-ibs}
This section presents insights from the literature and the Jakarta case study \citep{nespeca2020towards} that were used to formulate propositions and design the ABM for testing the propositions. 

\subsection{Mechanisms for the emergence of informational boundary spanners}
Insights regarding potential mechanisms for the emergence of IBSs were gained in the case study of Jakarta. In the words of an information management officer interviewed by \citep{nespeca2020towards}:

\begin{quote}
    "\textit{I was becoming a reference for everyone asking about mailing lists, who is working in a certain area, or what sort of maps are available. So that's the role that I have played}". UN-OCHA Information Management Officer. Collected in October 2018. 
\end{quote} 

This quote shows that an actor can emerge as an IBS not because of direct choice, but through an emergent process resulting from the collective choices and adjustments in the information collection preferences of a multitude of actors. This finding hints at the ability of actors to \textit{learn} the contacts that consistently provide relevant information and adjust their information collection preferences to match such contacts over time. This learning process constitutes a mechanism that occurs at the micro (agent) level, and can lead to the emergence of '\textit{information hub roles}' (here termed IBSs) at the macro level \citep[pp 9-10]{nespeca2020towards}. Such a mechanism is referred to as 'LearNing' or LN in the following. In this study, learning is compared to another mechanism in which actors collect information randomly among their contacts without developing information collection preferences (called 'Random  Collection' or RC).

Agents relying on LN develop a preference for collecting from contacts that share relevant information frequently. This significantly increases their chances of finding relevant information. For IBS candidates with inter-group contacts, these preferences tend to favor external contacts, who provide access to external information not readily available within the their own groups. Consequently, IBS candidates relying on LN are more likely to engage with sources outside their group boundaries, accessing new and relevant information more effectively than those using RC, who lack such targeted preferences. Therefore, IBS candidates influenced by LN are more likely to fulfill the informational boundary spanning function (Cf. Section measuring the emergence of IBSs) and emerge as effective IBSs compared to those that rely on RC. This has two consequences. First, a larger portion of IBSs candidates become emergent IBSs, resulting in a higher number of emergent IBSs with LN compared to RC. Second, as a whole, the IBSs that emerge are more effective in finding external information that is relevant and new for their groups with LN compared to RC. The following proposition is formulated. 

\vspace{2mm}
\noindent
\textbf{Proposition 1 (Learning VS Random Collection):}{When actors learn by adjusting their preferred information sources based on the past quality of information provided by such sources, a higher number of informational boundary spanners emerge that are more effective in retrieving external information than in the case in which actors collect information randomly among their contacts.}

\subsection{The effect of environmental uncertainty}

In disaster response, there are typically two types of events in a group's external environment that need to be detected by individuals belonging to the group to inform decision making, namely shocks and announcements \citep{nespeca2020towards, nespeca_methodology_2023}. Shocks are unexpected disruptive events such as cascading effects generated by infrastructural failures (e.g., blackouts), riots, or natural disasters. Announcements represent the release of information that is produced and consistently shared by particular groups or agencies such as flood early warnings \citep{watts_conceptualizing_2019, nespeca_information_2018}, evacuation orders \citep{adam_modelling_2017}, or needs assessments. 

In this study, uncertainty can be associated with the event itself  (e.g., concerning the nature and timing of the event) and the source  from which information regarding the event is released or made available (i.e. the information origin). Shocks and announcements differ in terms of their uncertainty in the source of information associated with them. In the case of shocks, it is not known when and where a shock will occur, and who will be affected and communicating about it. Consequently, the origin (or source) of information becomes uncertain. As such, shock-related information has an unstable origin. In contrast,  announcements are consistently generated from the same source within an information exchange network (e.g., weather forecasting agency in the case of early warnings, or the village leader in the case of evacuation orders). As such, while the time and type of announcement is still uncertain (it is not known when an early warning will be necessary), the source of announcement information is known. Therefore, announcement information is referred to as information with a stable origin. 

\begin{figure}[h!]
    \centering  
    \includegraphics[width=0.45 \textwidth]{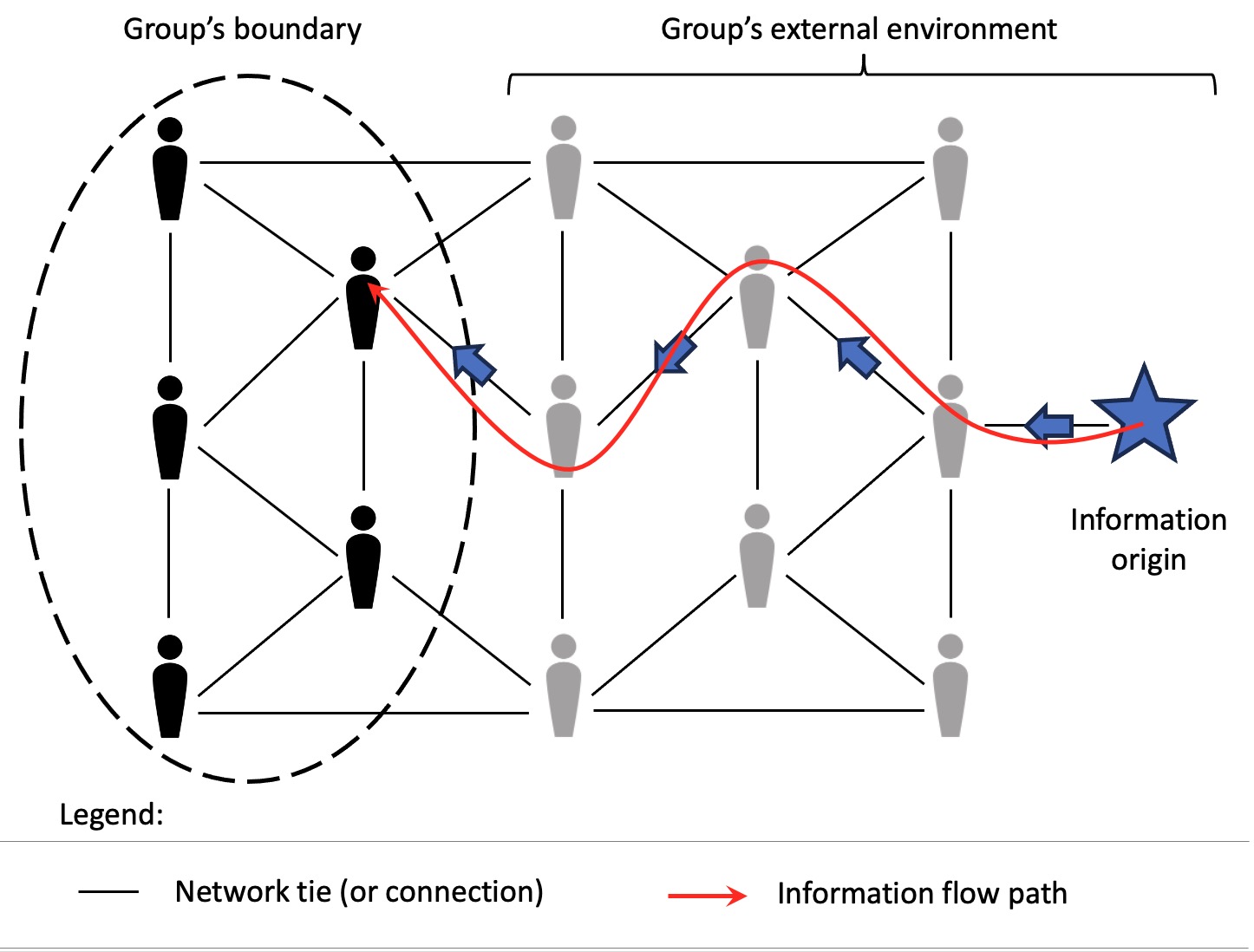}
    \caption{Information flow paths illustrating the spread of information from its origin (i.e. node in the network in which new information is introduced) through a series of information sharing activities carried out by the network nodes representing the actors.
    }
    \label{fig:stability}
\end{figure}

The combined effect of the stability in information origin and learning on the emergence of IBSs can be conceptualized from the perspective of \textit{information flow paths}. Information flow paths (or simply information flows) represent the contact-to-contact information exchanges through which information spreads within and across groups. These paths originate once information is created from unstructured data, and then routed to one or more of the actors belonging to the  groups considered (as shown in Figure \ref{fig:stability}).

The configuration of information flows and whether they reach groups depend on several factors. This study accounts for the following three: (1) information flows depend on the structure of the network through which information can be exchanged within and across the boundaries of different groups (as introduced by \citep{zagorecki_interorganizational_2009, altay_information_2013}). In this case, the network is considered to be  constant. (2) information flows are affected by the locations of the network, in which new external information is introduced i.e. the origin of the information. The origin of information can be stable (announcements), or unstable (shocks). (3) The paths depend on the information exchange behavior and preferences of the actors constituting the nodes in the network.

Whether learning impacts the emergence of IBSs and the ability of groups to retrieve external information is related to the stability of information origin. 
When the information origin is stable (announcements), the information flow paths originate in the same node. If that information is perceived to be relevant (as assumed for this study), the actors directly connected to the origin will learn and develop a preference towards choosing it as their source. Other actors who are only indirectly connected to the origin (i.e. via other contacts) are likely to develop a preference for the actors among their contacts that are the most directly connected to the origin (i.e. through the lowest number of ties), given they are more likely to consistently receive and share such information earlier than other contacts. Then, at each information exchange along the path, actors are likely to collect from the contacts that are the most directly connected with the information origin. As such, learning is expected to generate shorter information flow paths from the origin of information to the group that needs such information compared to the case in which actors collect information randomly and do not develop preferences.

Conversely, if information has an unstable origin (shocks) the source of information changes, and, thus, agents cannot learn which sources continually provide relevant information. The actors therefore will benefit less from developing information collection preferences. These considerations lead to the following proposition: 

\vspace{3mm}
\noindent
\textbf{Proposition 2 (Stability of information origin and learning):}
\textit{Learning leads to more effective inter-group information exchange if and only if relevant information is consistently generated by the same actors. If relevant information becomes available from continually changing and uncertain sources (unstable origin) learning has no effect in terms of improving inter-group information exchange. }

\subsection{The effect of environmental turbulence} 
Environmental turbulence is the frequency of environmental change \citep{duncan1972characteristics}. Specifically, turbulence in this study is defined as the frequency of occurrence of events (i.e., shocks, and announcements) representing changes in the environment. In order to detect and adapt to such changes, actors need to find information regarding these events. As such, the higher the level of turbulence the higher the volume of external information that agents need to manage and find. However, actors have a limited capacity to process information \citep{simon1955behavioral}. In the case of disasters, the volume and speed of information can be so prominent that actors reach the limit of their cognitive capabilities and become overloaded with information, which impairs their ability to retrieve and exchange the information needed \citep{van2016improving, comes2016cognitive}. As such, the authors of this study posit that, for higher levels of turbulence, the performance of the system decreases. Performance is measured as the amount of  information concerning environment-altering events that is collected or received by the groups that need such information. 

The number of actors that can explore the external environment was shown to play a crucial role in helping groups to detect environmental change  \citep{hazy_impact_2003}. In their study, the actors able to explore the external environment are those with inter-group ties. An increased number of inter-group ties entails a greater number of IBSs candidates who can emerge as IBSs and contribute to detect information concerning events marking environmental change. This is particularly relevant for high levels of turbulence, in which the high volume of information needs to be distributed among a higher number of IBSs to reduce the risk of information overload and enhance system's performance.

Additionally, LN is expected to increase the number of IBSs and improve their performance compared to random collection for high levels of turbulence. Considering that IBSs tend to process and exchange a high volume of information across different groups, they are likely to become overloaded with information. When the available IBSs are no longer able to provide relevant information because of information load, the actors requiring information from them will adapt their information sources through LN to find other actors able to provide the information they need. This may lead the actors to develop preferences towards IBSs candidates that are not yet as overloaded and that thus can support the exchange of information. As such, the adaptability introduced by learning is anticipated to enable actors to collectively distribute the volume of information across a higher number of IBSs, and thus improving overall system's performance compared to the case of RC. This performance increase is dependent on the agents' capacity to select from various IBS candidates resulting from a high number of inter-group ties, leading to the following proposition. 

\vspace{3mm}
\noindent
\textbf{Proposition 3 (Interplay of turbulence, inter-group ties redundancy, and learning):}
\textit{A higher number of ties among groups results in a higher number of emergent IBSs, enhancing a group's ability to retrieve external information even in highly turbulent environments. This enhanced ability to retrieve external information is particularly pronounced when actors rely on learning to adjust their information collection preferences.}

\section{Model Design}

Agent-based modeling (ABM) is the modeling paradigm of choice as it enables researchers to explore the collective consequences of individual behavior \citep{epstein1996growing}. 

\begin{figure*}[htp]
    \centering  
    \includegraphics[width=0.7\textwidth]{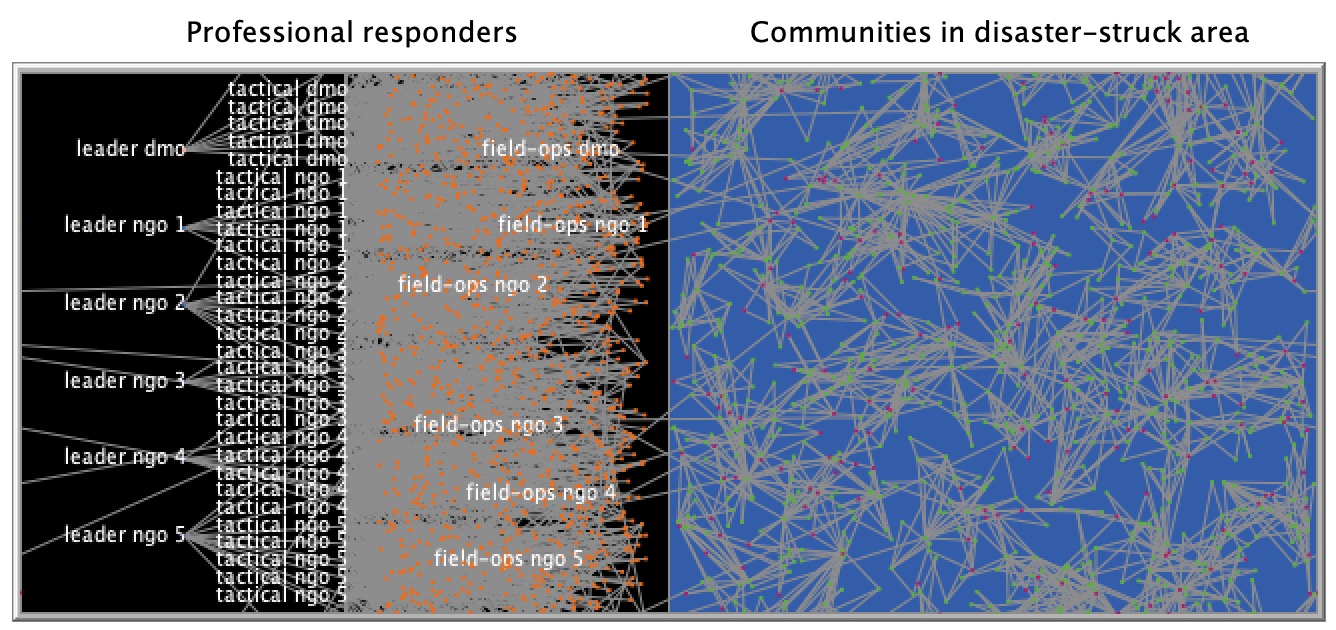}
    \caption{Graphical overview of the Agent-Based Model (ABM) developed for this study. The model involves two distinct groups, namely professional responders (on the left, in black) and communities (on the right, in blue) that exchange information. The grey lines represent the formal and informal ties used for information exchange. The ties that cross the border between professional responders and communities are the inter-group ties.}
    \label{fig:model_gui}
\end{figure*}

\begin{figure*}[htp]
    \centering  
    \includegraphics[width=0.85\textwidth]{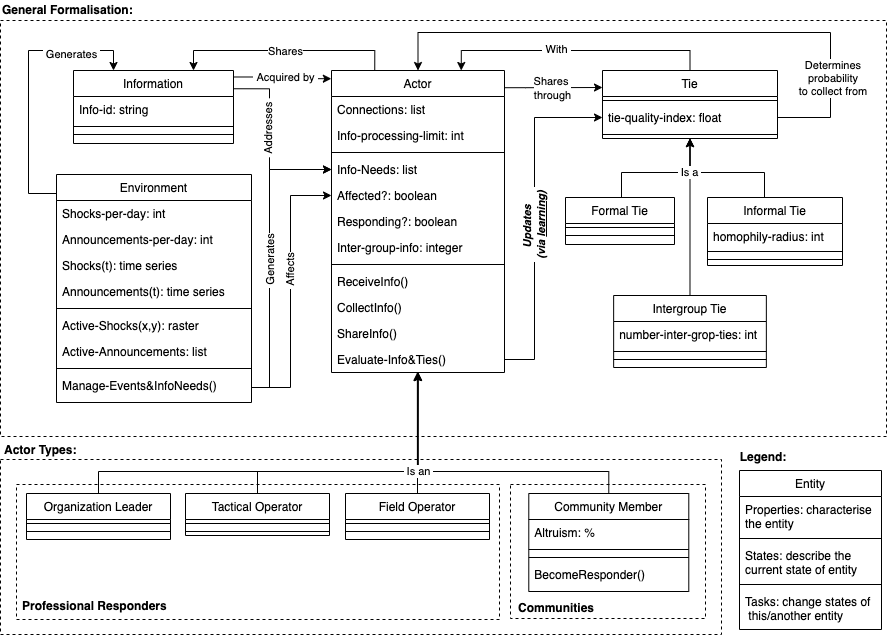}
    \caption{ Graphical description of entities, their properties, states, and tasks in the developed ABM. The actors and environment are agents in the ABM given they carry out tasks.}
    \label{fig:model_UML}
\end{figure*}

An ABM was developed based on the methodology and ABM proposed in \citep{nespeca_methodology_2023} and modified to include the learning mechanism found for the Jakarta case study in \citep{nespeca2020towards}. The goal of the model is to assess the validity of the method for measuring the emergence of IBSs and to study the mechanisms that influence the emergence of IBSs. A graphical overview of the results of this ABM is shown in Figure \ref{fig:model_gui}. Further, a conceptual diagram of the entities, states, and tasks included in the model is shown in Figure \ref{fig:model_UML}. Entities represent essential features of the model with their properties, states, and tasks as presented in the legend of Figure \ref{fig:model_UML}. 

Two main groups of agents are considered in the ABM, namely communities and professional response organizations or simply professionals (including governmental and non-governmental organizations). The system's performance is measured in terms of the percentage of external information needed by each of these two groups that is found by at least one of the actors belonging to the group. Each group has direct access to the information that the other group needs. Therefore, sharing information across group borders is vital. Specifically, communities have direct access to information about shocks, which represent the information needed by professionals. Conversely, professionals can send announcements, which represent the external information needed by communities. The exchange of information among the two groups is enabled through the emergence of IBSs. 

Information exchange within and across the two groups takes place via networks of informal and formal ties or contacts (Cf. Figure \ref{fig:model_gui}). Informal ties are obtained through preferential attachment algorithms based on  \citep{barabasi1999emergence}. Formal ties representing hierarchical structures are introduced within the professional response organizations according to three levels, namely strategic, tactical, and operational. Within such networks, IBS candidates (i.e., those that can potentially emerge as IBSs) are those that have at least one connection with an actor belonging to a group different from their own (cf. the ties connecting Professional Responders and Communities in Figure \ref{fig:model_gui}).

The following paragraphs discuss the key features introduced in the model to measure the emergence of IBSs and explore each of the propositions. 

\textit{Measuring the emergence of IBSs:} emergent IBSs are those that consistently provide \textit{new and external information} to a group thus playing a key role in fostering information exchange across the groups (cf. Section "\textit{Measuring the emergence of informational boundary spanners}"). To apply such a definition, it is crucial to measure each agent's contribution to information sharing across groups. This measurement is facilitated by the introduction of the '\textit{number of informational boundary-spanning Function Executions}' or FEs, where an actor agent increases its FE count each time it provides new and external information to any group that requires it. The number of FEs accumulated by agents are then used to study which of the IBS candidates emerge as IBSs. 

\textit{Understanding emergence:} the first proposition focuses on studying the emergence and effectiveness of IBSs by comparing random information collection (RC) with learning (LN). RC and LN are characterized by different information collection preferences. Such preferences are modeled as non-uniform distributions in the probability that agents have to collect from each of their contacts, denoting that some of the agents' contacts are preferred compared to others. For RC, the information collection probabilities assigned by an agent to its contacts are equally divided among all contacts, indicating that the agent has no preferences (uniform distribution). For LN, each agent develops and adjusts its information collection preferences over time (possibly leading to a non-uniform distribution). Specifically, a reinforcement learning algorithm is used to adjust preferences over time for each agent based on the relevance of the information provided by each of the agent's contacts from the beginning of the simulation up to the current time step. The adopted reinforcement learning algorithm is Q-learning (see Appendix A).

The second proposition considers the interplay of learning and uncertainty. To test the impact of stability in information origin, announcements (with a stable information origin) and shocks (with an unstable information origin) are introduced in the model. The degree of uncertainty is determined by the parameters \textit{shocks per day} and \textit{announcements per day}. For instance, if \textit{shocks per day} is set to 0 and \textit{announcements per day} is set to a value greater than 0,  environmental uncertainty is considered to be low, as only stable-origin announcements are considered. 

The third proposition focuses on studying the combined effect of learning, turbulence, and number of inter-group ties on the emergence and performance of IBSs. Varying levels of turbulence are captured by the different frequencies in the number of shocks and announcements (parameters \textit{shocks per day} and \textit{announcements per day}) considered. The number of ties across communities and professional responders are specified by the parameter \textit{number of inter-group ties}. The inter-group ties are introduced in the model with a preferential attachment algorithm modified to choose pairs of agents belonging to different groups with a likelihood that depends on their current degree (i.e. number of ties). Bounded rationality is captured by limiting the amount of information that actors can process, share, and collect to 3 pieces of information within each simulation step (set the 'info. processing limit' parameter in the model).

Regarding the temporal and spatial scales considered in this model, the time step of simulation is 10 minutes, while the duration of the simulation is 4 days. Spatial scale is considered at an abstract level in this model as the actors exchange information through networks. As such, a specific spatial scale is not assigned in this case.

\section{Methods}
This section describes the way the model parameters were set and the experiments that were run to test the propositions.


\subsubsection{Model Parametrization}
 Three parameters were constant for all experiments namely:  duration  of a simulation, the learning rate, and the information processing limit. The duration of each simulation was set to 4 days to simulate information exchange in the early stages of disaster response, a phase in which volatility is especially pronounced and rapid inter-group information exchange and coordination are particularly crucial  \citep{kreiss2010early, bode2017stages}. The learning rate was set to a relatively low value (assumed to be 0.1) as the time step in the simulation is small (10 minutes) compared to the chosen duration of the simulation, requiring a smaller learning rate to compensate for the high frequency with which information exchange preferences are updated (up to every 10 minutes). Finally, the information processing limit was set to a value of 3, representing humans' limited ability to process information per unit of time (in this case every 10 minutes) \citep{simon1955behavioral}. 


\subsubsection{Experimental design}
In total, four batches of experiments were designed to study (a) the method for measuring the emergence of IBSs and (b) each of the 3 propositions as shown in Table \ref{tab:experiments}. Experiment 0 focuses on testing the approach for capturing and measuring the emergence of IBSs. The simulations are run for varying numbers of inter-group ties. 
Experiment 1 focuses on simulating the impact of the two different information collection mechanisms to study Proposition 1. Next, Experiment 2 studies Proposition 2 and thus investigates the interplay between learning and varying levels of uncertainty (stability in information origin) by combining different levels of \textit{shocks per day} and \textit{announcements per day}. Finally, Experiment 3 explores Proposition 3 and thus focuses on the interplay between varying levels of environmental turbulence (established with the parameters \textit{shocks per day} and \textit{announcements per day}), learning (LN VS RC), and the number of inter-group ties (established with the homonym parameter). These experiments are run through the service provided by the \citet{DHPC2022}.

\begingroup

\renewcommand{\arraystretch}{1.5} 

\begin{table*}[h!]
	\centering
	\begin{tabular}{p{0.15\textwidth}p{0.17\textwidth}p{0.17\textwidth}p{0.17\textwidth}p{0.17\textwidth}}
	\toprule
 	\textit{Parameters} & Experiment 0 \newline (method): \newline Measuring  emergent IBSs & Experiment 1 \newline (Proposition 1): \newline Comparing LN and RC  & Experiment 2 \newline (Proposition 2):  \newline Effect of Environ.  Uncertainty  & Experiment 3 \newline (Proposition 3):  \newline Effect of Environ. Turbulence \\ 
	\midrule

        \textit{Info. Collection Mechanism} & RC & LN, RC & LN, RC  & LN, RC  \\
    	
        \textit{shocks per day} & 10 & 10 & 0, 10, 20 & 1$^*$, 5$^*$, 10$^*$, 15$^*$, 20$^*$ \\  
        \textit{announcements per day} & 10 & 10 & 0, 10, 20 & 1$^*$, 5$^*$, 10$^*$, 15$^*$, 20$^*$ \\
        
        \textit{number of inter-group ties}  & 1, 2, 5, 10, 20, 30 & 20 & 20  & 1, 2, 5, 10, 20, 30 \\
    
        \textit{duration of the simulation (days)}  & 4 & 4 & 4 & 4 \\
    
        \textit{learning rate}  & N.A. & 0.1 & 0.1 & 0.1 \\
    
        \textit{info. processing limit}  & 3 & 3 & 3 & 3 \\
    
        \textit{number of repetitions}  & 80 & 40 & 40 & 20 \\
    
        \textit{total simulations}  & 120 & 80 & 720 & 1200 \\
    
	\bottomrule	
	
	\end{tabular} 
	
	\caption{Parameters setting for the simulation experiments aimed at testing each of  the propositions. The experiments were full factorial with the exception of the values marked with an asterix '$^*$'. The use of $^*$ for experiment 3 indicates that for each simulation in the experiment an equal value of shocks per day and announcements per day is considered (e.g. 10 shocks per day and 10 announcements per day) rather then their full factorial combination. LN = LearNing, RC = Random Collection.}
	
	\label{tab:experiments}	
\end{table*}
\endgroup

\section{Results}
This section illustrates the results of the experiments shown in Table \ref{tab:experiments}. 

\subsection{Experiment 0: Measuring the emergence of informational boundary spanners}

\begin{figure}[h!]
    \includegraphics[width = 0.44\textwidth]{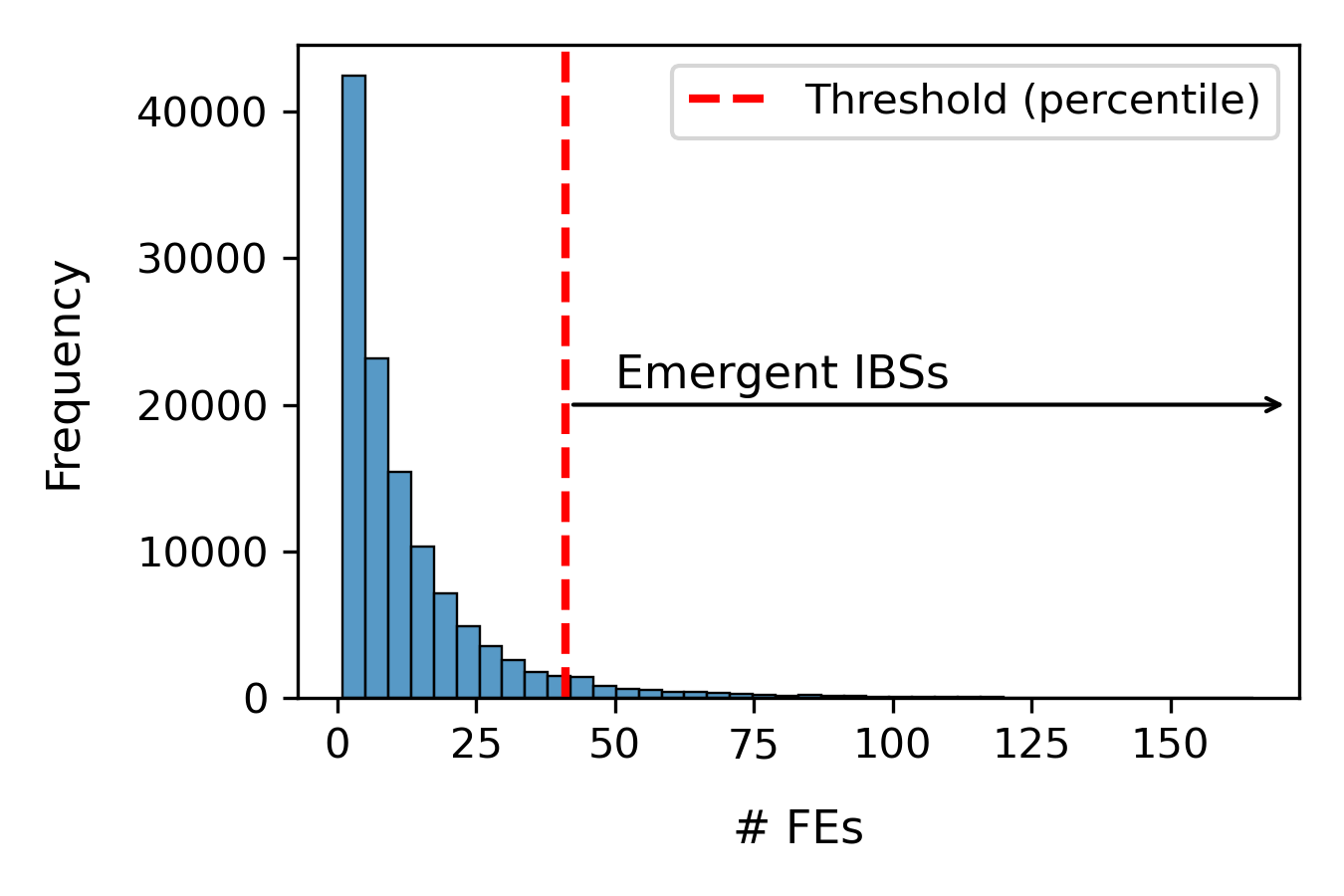}
    \caption{Frequency of occurrence of \# FEs (number of informational boundary-spanning Function Executions) obtained by the IBS candidates, and the threshold used to capture the emergent IBSs.}
    \label{fig:exp-0_individual}
\end{figure}

Figure \ref{fig:exp-0_individual} illustrates the distribution of the number of instances where agents with inter-group ties (IBS candidates) effectively contribute new and relevant external information to the groups that need it, earning them informational boundary spanning Function Executions (FEs). 

The distribution reveals that several IBS candidates attain relatively few FEs, while only a few candidates achieve a high number of FEs. This finding illustrates that a few IBS candidates contribute to a great extent to inter-group information exchange by providing new and relevant information to the groups who need it, while many other candidates do not contribute significantly. The candidates that contribute significantly to inter-group information exchange qualify as emergent IBSs. One approach to identify these candidates is to assume that those that achieve a number of FEs above a given threshold contribute effectively also at the inter-group level and thus qualify as emergent IBSs (cf. the red vertical line in Figure \ref{fig:exp-0_individual}). Such threshold or thresholds should be high enough to avoid capturing those IBS candidates that obtain only a few FEs (and as such do not considerably improve inter-group information exchange), but also not too high to avoid not capturing agents that may not have the highest FEs but still contribute significantly to inter-group information exchange. However, it is unclear how to select a threshold that satisfies these conditions. 

\begin{figure}[h!]
    \includegraphics[width = 0.42\textwidth]{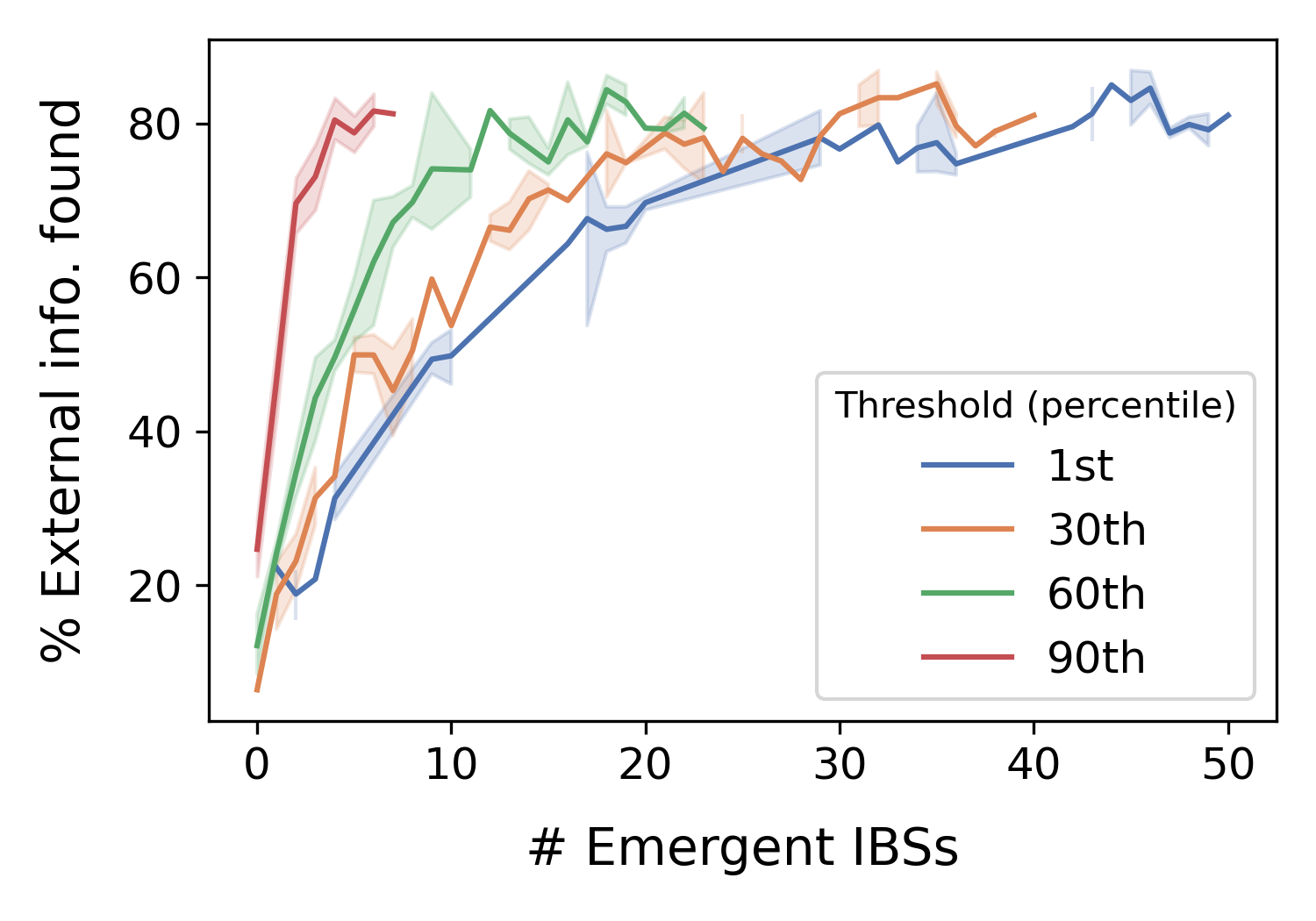}
    \caption{Relationship between the number of IBSs that emerged and the \% of external information found by the groups who need it for different values of the threshold set as a percentile in the distribution of FEs from Figure \ref{fig:exp-0_individual}.}
    \label{fig:exp-0_threshold_test}
\end{figure}

To investigate the effects of the choice of this threshold, Experiment 0 focuses on studying the implications of adopting different thresholds, namely the 1st, 10th, 20th, 30th, 40th, 50th, 60th, 70th, 80th, and 90th percentiles in the frequency distribution shown in Figure \ref{fig:exp-0_individual}. For each of these thresholds, the number of IBSs emerged is plotted against their effectiveness at the inter-group level, measured as the \% of external information needed found by the groups. Figure \ref{fig:exp-0_threshold_test} illustrates the results for selected thresholds (i.e. the 1st, 30th, 60th, 90th percentiles) to enhance clarity of representation. These results show that for all considered thresholds a higher number of emergent IBSs corresponds to a higher performance. Further, the number of emergent IBSs  grows considerably when increasing the threshold. This figure, however, does not consider the number of IBS candidates available, which represents the maximum number of IBS that can emerge. As such, a more in depth analysis is required that considers not only the threshold adopted, but also the number of IBS candidates available.

\begin{figure*}[h!]
    \centering  
    \includegraphics[width = \textwidth]{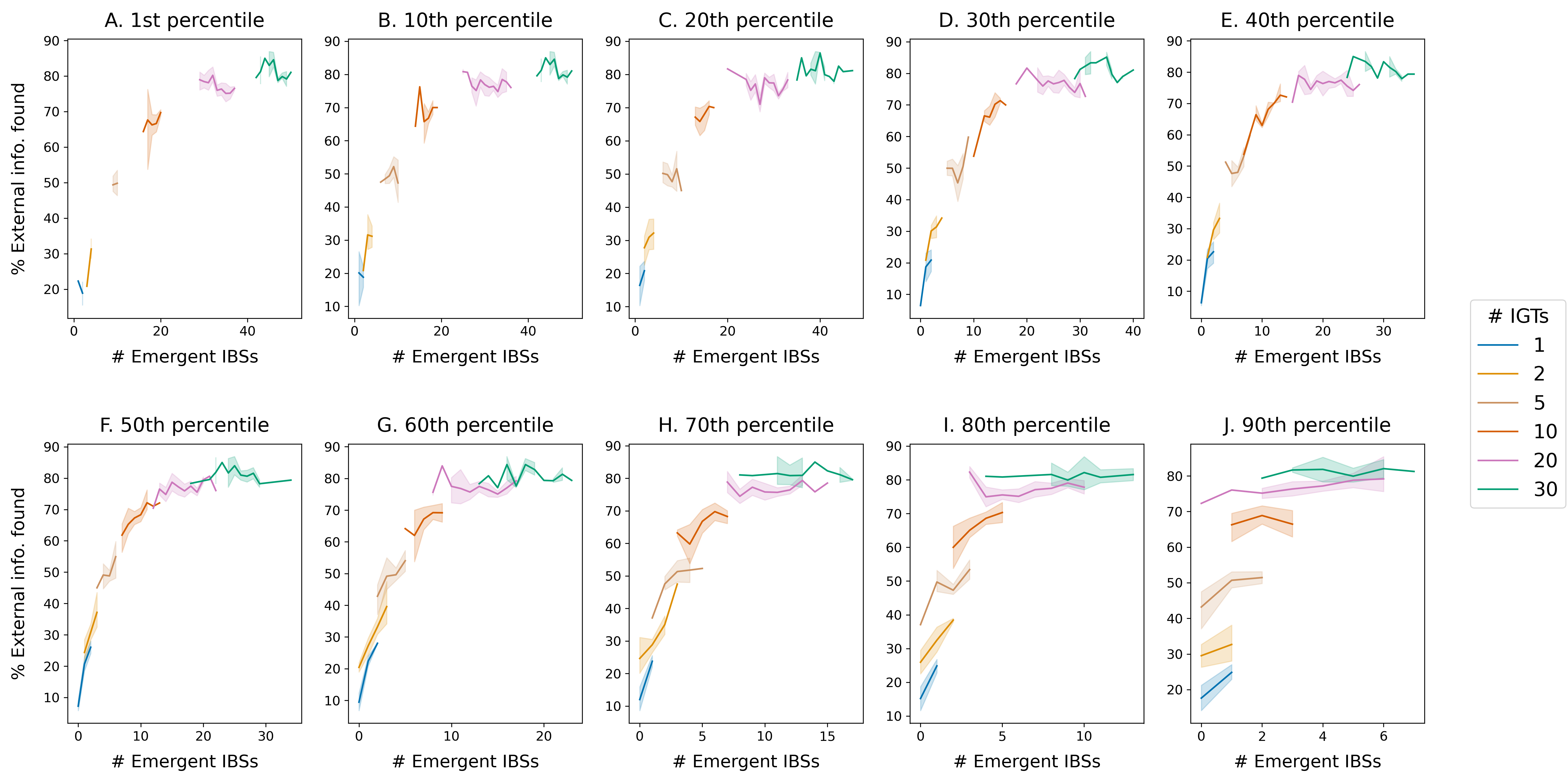}
    \caption{Number of IBSs that emerged for each of the thresholds considered in Figure B and for different numbers of Inter-group Ties (\# IGTs).}
    \label{fig:exp-0_IGTs}
\end{figure*}

To analyze emerging IBSs across varying numbers of IBS candidates and threshold settings, it is key to consider that the number of IBSs candidates depends on the number of Inter-Group Ties (\# IGTs) available among groups. Precisely, the number of IBS candidates is twice \# IGTs given that each connection ties two agents who can both emerge as IBSs. As such, the analysis depicted in Figure \ref{fig:exp-0_threshold_test} is expanded to consider not only different thresholds (in this case all of those between the 1st and 90th percentiles), but also different \# IGTs. Figures \ref{fig:exp-0_IGTs}.A to \ref{fig:exp-0_IGTs}.J depict the results.  A comparison of these figures reveals that, at low thresholds, as depicted in Figures \ref{fig:exp-0_IGTs}.A to C, nearly all IBS candidates transition into emergent IBSs. For instance, with 10 inter-group ties (meaning 20 IBS candidates), the emerging number of IBSs tend to be around 20, underscoring the high conversion rate of IBS candidates into actual IBSs (cf. Figure \ref{fig:exp-0_IGTs}.A, B, and C). Furthermore, the simulation results cluster in specific areas of the plot when holding the number of inter-group ties constant. This clustering suggests minimal variability in both the number of emerging IBSs and the overall system performance. Such lacking variability shows the little sensitivity in capturing emergent IBSs provided by low thresholds. Conversely, for high thresholds such as the 90th percentile (Figure \ref{fig:exp-0_IGTs}.D), there is significant variability in the number of emergent IBSs, yet the percentage of external information remains relatively constant, as shown by the horizontal lines. This shows that the emergent IBSs captured with high thresholds do not contribute significantly to effective inter-group information exchange. Thresholds between these extremes, like those between the 30th and 80th percentiles (Figures \ref{fig:exp-0_IGTs}.D to \ref{fig:exp-0_IGTs}.I), show variability in the number of IBSs emerged. Further, for an increase in the \#IBSs, the \% of external information found also increases, showing how the additional emergent IBSs captured positively contribute to fostering inter-group information exchange. One exception to such an increase is observed for the 80th percentile and 20 IGTs when the number of emergent IBSs found is 3. This is most likely an outlier given that the plot does not show a confidence interval around the value. 

These findings illustrate that the method proposed in this paper captures the number of emergent IBSs that significantly contribute to providing new and external information to groups, thereby supporting effective inter-group information exchange. 

However, for this method to be effective, thresholds need to be carefully chosen according to two criteria. First, adequate thresholds are not too low, thus presenting sensitivity to the emergence of varying numbers of IBSs across different simulations. Second, such thresholds are not too high, meaning that when the number of IBSs captured through the threshold increases, also their effectiveness in fostering inter-group information exchange increases.  

These two criteria are sufficient to find multiple adequate thresholds rather than a single one (in this case those between the 30th and 80th percentiles), leaving an open question as to whether adopting different adequate thresholds will provide different results when studying the emergence of IBSs. As such, rather than simply choosing one of the adequate thresholds, it is key to use multiple thresholds and assess the consistency of the findings obtained. The remaining experiments (1 to 3) are analyzed with six thresholds, namely the 30th, 40th, 50th, 60th, 70th, and 80th percentiles. For brevity, the following sections show the results for the 60th percentile. However, a comparison of the IBSs emergence results obtained with the six percentiles chosen are presented in Appendix B. 

\subsection{Experiment 1: Learning VS Random Collection}
Proposition 1 is composed of two parts. The first part indicates that when actors learn and adjust the information collection preferences based on the quality of information provided by their contacts (LeaRning mechanism or LN) this leads to the emergence of more IBSs compared to the case in which actors collect information randomly (Random Collection mechanism or RC). The second element of the proposition states that when actors learn (LN), this leads to the emergence of IBSs that are more effective in fostering inter-group information exchange compared to random collection (RC). In the following these two parts are assessed against the results of experiment 1.

\begin{figure}[h!]
    \includegraphics[width=0.48\textwidth]{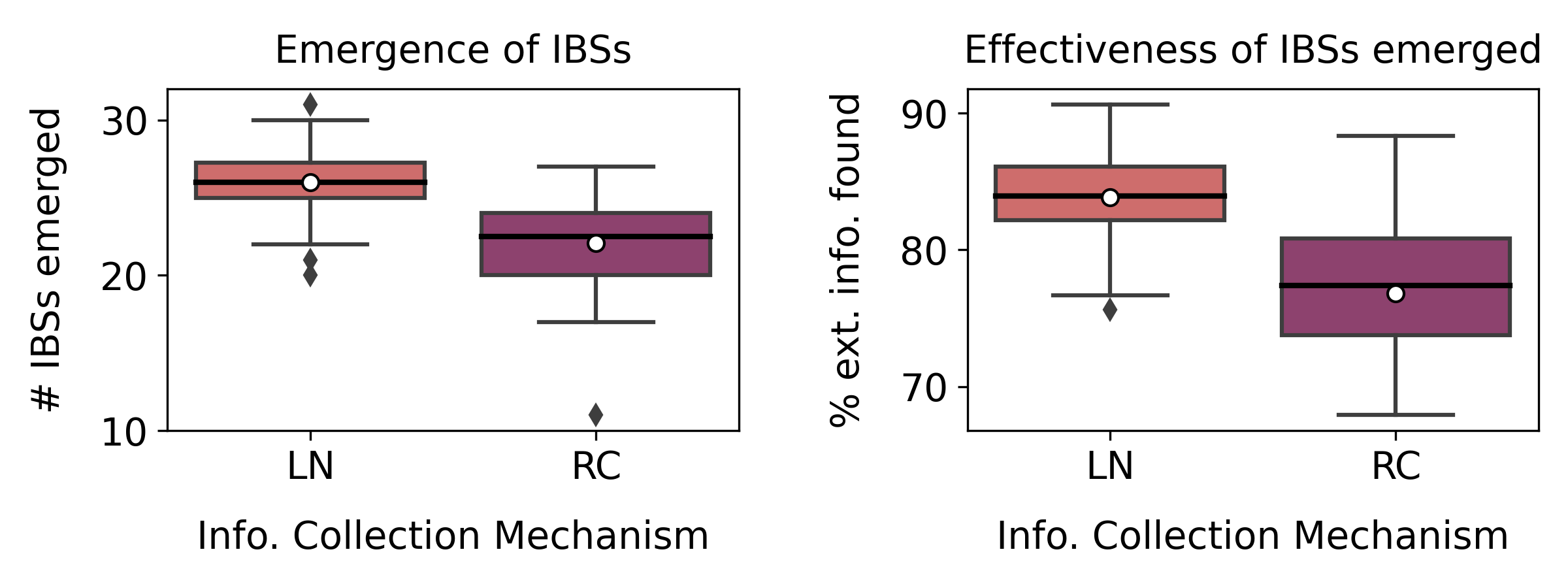}
    \caption{Results of Experiment 1: Effect of the information collection mechanisms Random Collection (RC) and LearNing (LN) on the \# IBSs emerged (left) and the \% of external information found by the groups that need it (right). The white dots represent the averages.  
    }
    \label{fig:effect-of-learning-ibs}
\end{figure}

Figure \ref{fig:effect-of-learning-ibs} shows the results of experiment 1 including the emergence of IBSs (left) and their performance (right) respectively for the information collection mechanisms LN and RC. 

\begin{figure*}[t!]
    \centering  
    \includegraphics[width=0.65\textwidth]{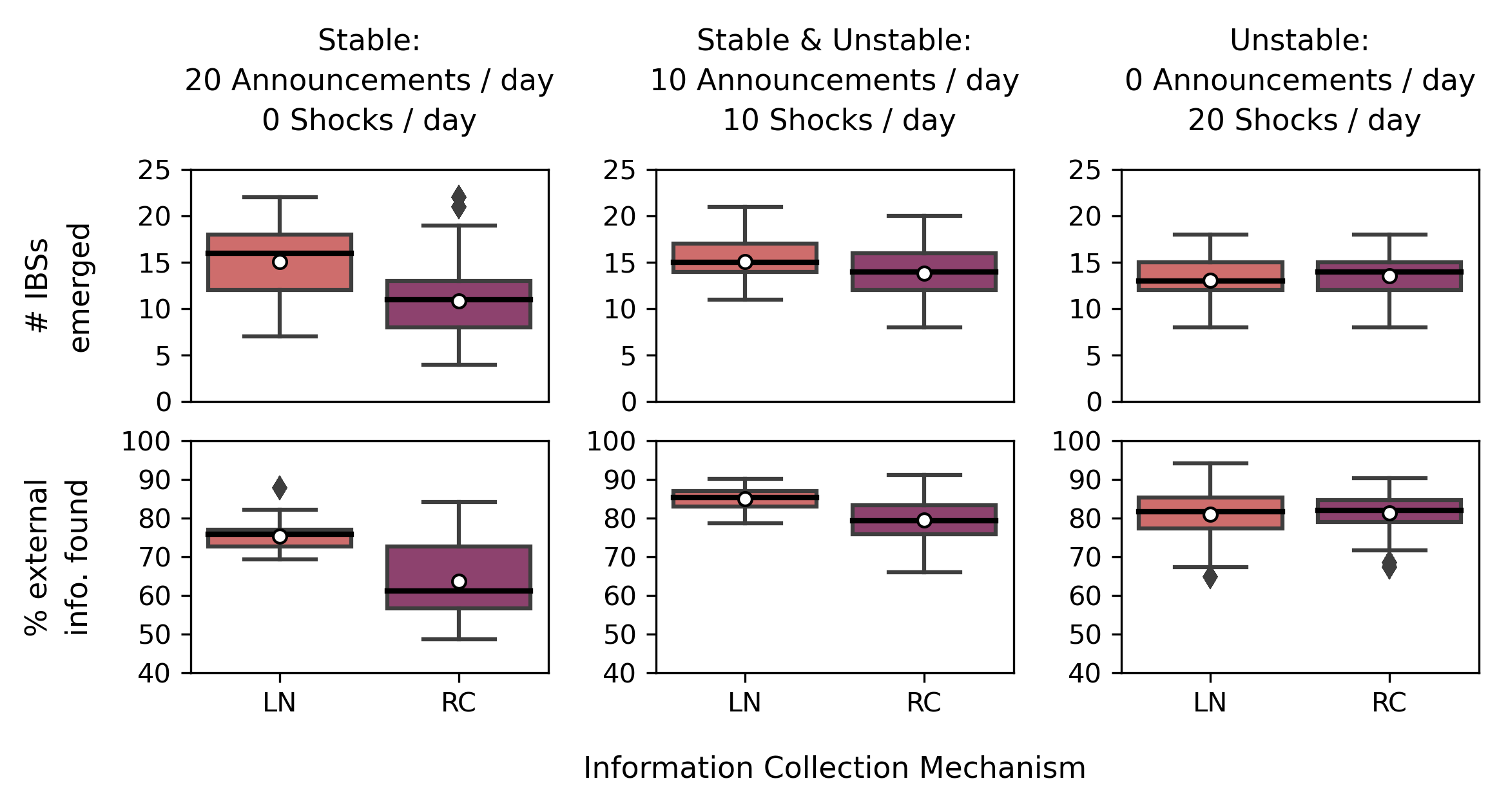}
    \caption{Results of Experiment 2: Interplay between the information collection mechanisms Random Collection (RC) and LearNing (LN), and varying environmental uncertainty. The effect of such an interplay is assessed for (a) the emergence of  Informational Boundary Spanners (IBSs) (upper row) and on (b) the effectiveness of such IBSs in enabling groups to retrieve external and relevant information (lower row).  The varying environmental uncertainty is characterized by stable (20 announcements per day - left column), mixed stable and unstable (10 announcements per day and 10 shocks per day - middle column), and unstable (20 shocks per day - right column) origins of external information.The white dots represent the averages.}
    \label{fig:exp-2}
\end{figure*}

First, Figure \ref{fig:effect-of-learning-ibs} (left) illustrates how, compared to RC, LN results in an average increase of about three emergent IBSs. The 5th, 25th, 75th, and 95th percentiles and the median are also shifted towards higher values in the case of LN compared to RC. The interquartile range of the number of IBSs emerged is larger for LN than in the case of RC, showing that learning also increases the variability of the results. However, the 5th percentile is closer to the median in the case of LN compared to RC, illustrating how, despite its higher variability, the distribution is skewed towards higher values of \# IBSs emerging in the case of LN compared to RC. These results supports the first part of Proposition 1 by showing that the number of IBSs emerged tends to increase with LN compared to RC. 

Second, according to Figure \ref{fig:effect-of-learning-ibs} (right) the effectiveness of IBSs measured as the percentage of external information retrieved by each group increases roughly by 7.5\% on average with LN compared to RC. The 5th, 25th, 75th, and 95th percentiles, and the median shift towards higher values in the case of LN compared to RC. Additionally, the interquartile range for LN is considerably reduced compared to RC, illustrating how, in combination with the higher median, the retrieval of external information is more likely to be higher in the case of LN compared to the case of RC. These results support the second part of proposition 1 by showing that when agents learn (LN) the IBSs that emerge are able to find and provide higher percentages of the external information needed by groups compared to the case in which information is collected randomly (RC). In sum, both parts of Proposition 1 are supported by the results. 

\subsection{Experiment 2: The effect of environmental uncertainty (stability of information origin)}
Figure \ref{fig:exp-2} shows the results of Experiment 2 regarding the emergence of IBSs (upper row) and their performance (lower row) for both LN and RC. These results include different combinations of frequencies in shocks and announcements, namely 20 announcements per day (left column), 10 shocks per day and 10 announcements per day (middle column), and 20 shocks per day (right column). In all three cases the number of events (shock, announcements, or combinations of them) amount to a total of 20 for each day of simulation to ensure comparability among them.  

The results show that, in the case of announcements (Cf. left column of Figure \ref{fig:exp-2}), LN increases the \# IBSs emerged (by roughly 7 IBSs) and their performance (by around 15\%) compared to RC. 

When announcements and shocks are combined, as illustrated in the middle column of Figure \ref{fig:exp-2}, LN leads to an average emergence of 2 additional IBSs and a performance increase of 7\% over RC. However, these gains are modest when compared to scenarios solely involving announcements. For announcements only (see left column of Figure \ref{fig:exp-2}), there was a notable rise of 7 emergent IBSs and a 15\% performance improvement. 

Next, in the case of shocks (Cf. right column of Figure \ref{fig:exp-2}), the number of IBSs that emerge and their performance does not change significantly with LN or RC. 

These findings support Proposition 2 by showing that the effect of learning depends on stability of the information origin. If the information origin is stable (announcements), then learning leads to an increase in the number of IBSs and in their effectiveness in improving inter-group information exchange. Conversely, if the source of information is unstable (shocks), learning has little effect. 


\subsection{Experiment 3: The effect of environmental turbulence}
\begin{figure}[h!]
    \includegraphics[width=0.48\textwidth]{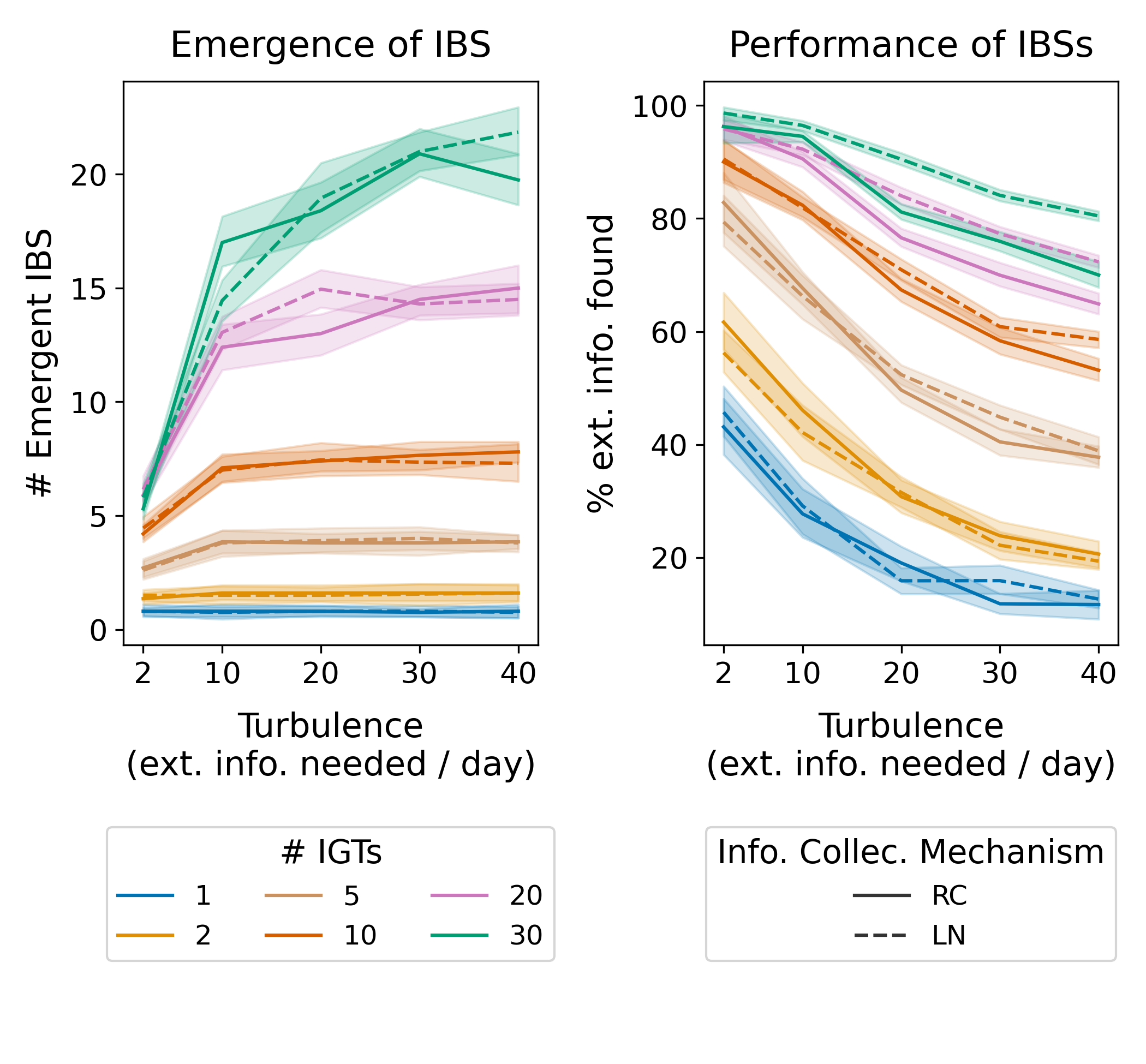}
    \caption{Results of experiment 3: Effect of increasing environmental turbulence on the number of  Informational Boundary Spanners (IBSs) emerged (left) and their collective performance in exchanging external relevant information across groups (right) for different numbers of Inter-group Ties (\# IGTs) between communities and professionals, and for the two information collection mechanisms LN (LearNing) and RC (Random Collection). 
    }
    \label{fig:exp-3}
\end{figure}

This experiment focused on the emergence of IBSs and their effectiveness under different levels of turbulence and numbers of Inter-group Ties (\# IGTs). The level of turbulence consists of the frequency of disruptive events (shocks and announcements) occurring every day of simulation and generating external information needs for the groups (measured as external information needed per day). In other words, the turbulence level is set as the sum of shocks per day and announcements per day. An equal number of shocks and announcements per day is considered in all experiments. As such, a Turbulence level of ten corresponds to five announcements per day plus five shocks per day. The results are shown in Figure \ref{fig:exp-3}. 

Figure \ref{fig:exp-3} on the left shows that for higher turbulence a higher number of IBSs emerges. The number of IBSs that emerge grows with the \# IGTs, and it does not change significantly with the information collection mechanism considered (LN or RC).  

Figure \ref{fig:exp-3} on the right shows that for higher levels of turbulence (and corresponding increasing information needs), the share of the external information that is retrieved decreases. However, a higher number of inter-group ties increases the performance of the system at all levels of turbulence. Further, such performance increases with LN compared to RC for high levels of environmental turbulence (10 or more events per day) if the number of inter-group ties is also high (20 or more IGTs).  

In sum, these findings support proposition 3 for two reasons. First, independently from the information collection mechanism, more inter-group ties are associated with the emergence of more IBSs that more effectively convey information across groups. Second, for high numbers of IGTs and levels of environmental turbulence, the performance of the IBSs emerged increases with LN compared to RC.

\section{Discussion}
This section discusses the implications of the findings, and suggests directions for future research.

\subsection{Measuring emergent IBSs}
This study introduced a method for measuring the emergence of IBSs. Compared to previous work that measures the emergence of informational boundary spanning at the level of a group or groups \citep{ancona1992bridging, hansen1999search, cummings2004work, zagorecki_interorganizational_2009, marrone2010team, van_osch_towards_2016}, the method introduced here directly measures the individual agents that emerge as IBSs. Emergence of IBSs can thus be tracked in greater detail, and the heterogeneous characteristics of those that emerge as IBSs can be observed at the micro level. Further, the interplay between the individual, group, and inter-group levels and its impact on the emergence of effective IBSs can be studied through this method. These aspects are crucial to inform the design of effective information management strategies that foster the emergence of IBSs, enhance inter-group information exchange, and support coordination and collective intelligence \citep{levina2005emergence}. 

To qualify as IBSs, the IBSs candidates need to fulfill a minimum "number of informational boundary spanning Function Executions" or FEs (i.e. a minimum threshold) that enables them to significantly improve information exchange among groups. To ensure the correctness of this method, thresholds must be carefully selected based on two criteria: they should not be too low to maintain sensitivity to varying occurrences of IBSs in different simulations, and not too high to ensure that as more IBSs are detected, their collective ability to enhance inter-group information exchange also increases. Here, inter-group information exchange is measured as the percentage of the total external information needed found by the groups. 

Experiment 0 showed that this method enables to measure emergent IBSs by individuating those actors that effectively contribute to inter-group information exchange. Further, multiple adequate thresholds were found to satisfy the two criteria mentioned above. To clarify whether adopting different adequate thresholds lead to different results when studying the emergence of IBSs, the results of Experiments 1 to 3 were analyzed and compared with six adequate thresholds: the 30th, 40th, 50th, 60th, 70th, and 80th percentiles of the FE distribution for IBS candidates. The results of this comparison in Appendix B show that adopting different adequate thresholds does not change the conclusions of the experiments as Propositions 1 to 3 remain supported by the findings independently of the thresholds considered. However, Appendix B also shows that it is still essential to consider a wide range of adequate thresholds when studying the emergence of IBSs. Specifically, a discrepancy was found in Experiment 2 in the results observed with the minimum and maximum thresholds (respectively, the 30th and 80th percentiles). These results were considered as outliers and thus discarded given their extreme value, inconsistency in the effects found, and the fact that the majority of the other thresholds indicated consistent results (cf. Appendix B). 
This process illustrates the importance of considering multiple thresholds and carefully analyzing any discrepancies that may arise. Such an analysis allows one to determine whether discrepancies indicate inconsistencies in the findings that require reconsideration or modification of the conclusions, or if they simply represent outliers resulting from the adoption of extreme thresholds (i.e., too low or too high). 

In sum, this method enables the study of the emergence of IBSs. To ensure the method's correctness, it is first crucial to select thresholds that are neither too low nor too high, thereby effectively capturing emergent IBSs that contribute to inter-group information exchange. Second, a wide range of adequate thresholds should be considered to test consistency across results on the emergence of IBSs and to assess whether any discrepancies are outliers or if the conclusions need to be reconsidered and modified in light of such discrepancies. 

\subsection{Understanding emergent IBSs}
First, this paper introduces a novel agent-based modeling (ABM) framework that enables to study the emergence of IBSs in volatile environments. This ABM simulates inter-group information exchange and outputs the number of times IBS candidates fulfill the informational boundary spanning function at the micro level, as well as their overall effectiveness. This provides the basis to identify those that emerge as IBSs through the proposed method and to systematically test mechanisms for their emergence in volatile conditions. Compared to previous ABM frameworks introduced to study inter-group information exchange and boundary spanning in volatile conditions \citep{hazy_impact_2003, zagorecki_interorganizational_2009}, this framework measures the emergence of IBSs rather than imposing a predefined number of IBSs a priori \citep{hazy_impact_2003}, and captures this emergence at the micro or individual level rather than solely at the macro level \citep{zagorecki_interorganizational_2009}.

Second, this study furthers the understanding of learning for the emergence of IBSs. Learning entails that each agent develops information collection preferences over time based on the past quality of information provided by the agent's contacts. This mechanism was compared to the case in which actors collect information randomly. As posited in Proposition 1 and supported by the results of Experiment 1, learning leads to the emergence of more IBSs as compared to random information collection, and these IBSs contribute to more effective inter-group information exchange. Learning therefore is a micro-macro mechanism as learning at the micro level leads to the emergence of IBSs and effective inter-group information exchange at the macro level. This finding confirms the results of the study by \citet{levina2005emergence} and extends it to volatile environments by showing that effective boundary spanners emerge through a process resulting from the decentralized interactions among actors belonging to different groups. It also adds to the work by \citet{marrone2010team} by showing \textit{how} micro level antecedents such as learning and the availability of connections with other groups can lead to macro level outcomes through an emergent process. 

Third, this study highlights the necessity to consider environmental volatility, i.e. the uncertainty and turbulence of the environment, for studying the emergence of IBSs. As posited in Proposition 2 and corroborated by the findings from Experiment 2, the effect of learning on the emergence and efficiency of IBSs is contingent on environmental uncertainty and specifically on the stability of the information source or origin. If the information origin is stable, as in the case of announcements, the number of IBSs and their performance in enhancing inter-group information exchange increases when the agents adopt learning. In contrast, an unstable information source, as in the case of shocks, renders learning ineffective. This difference can be explained by the differences of information origin and network structures: while announcements are constantly originating from the same location of the network, shocks originate from  random locations in the disaster-affected area of the model. Under learning, announcements  propagate through the network following increasingly strong preferential channels. In contrast, the way shocks propagate continually changes depending on where the shock occurred and which nodes/actors in the network find this information and share it with others. These results explain why learning has little effect on shocks or more generally information of unstable origin: when information can come from ever-changing locations of the network, developing information collection preferences for the contacts that provided the most relevant information in the past has the same effect as collecting information randomly given that none of the agents tends to consistently provide relevant information due to the instability of the information origin. 

Proposition 3 is supported by the findings of Experiment 3  and suggests that an increased presence of inter-group ties leads to the emergence of a higher number of IBSs that, as a collective, can more effectively facilitate the exchange of information between groups. This effect occurs even for high turbulence and when considering the actors' limited information processing and sharing capability. Further, for high levels of inter-group ties and environmental turbulence, learning leads to higher performance than in the case of random information collection. Such an improvement is negligible for low numbers of ties and becomes evident for 20 and 30 inter-group ties. This pattern can be attributed to the agents' collective capacity to discern through learning the most effective IBS candidates for relaying external information (e.g., those that are exposed to less information load) as they change over time. Further, the fact that learning has an effect only for 20 or more inter-group connections illustrates that a sufficiently extensive network of inter-group ties is crucial for enabling agents to choose from various IBS candidates through learning.  

\subsection{Implications for Collective Intelligence}
This study contributes to the collective intelligence literature by illustrating \textit{how} a cognitive process at the individual level (i.e., learning \citep{sternberg1982conceptions}) can support the collective selection of actors (the IBSs) that effectively convey information across multiple groups to support their coordination \citep{argote1982input,  wittenbaum2002coordination}. This can be considered to be collective intelligent behaviour given that the groups are able to select actors that are more effective than others at carrying out particular tasks or activities - in this case fostering inter-group information exchange - thus possibly improving the system's performance (groups' retrieval of the external information needed for decision making and coordination) \citep{malone2022handbook}. 

Further, this study illustrates that this collective intelligent behaviour is contingent upon the characteristics of the environments in which groups operate \citep{lawrence1967organization, duncan1979right} and specifically its volatility (characterized by uncertainty and turbulence). When environmental uncertainty is very high and, as such, information presents unstable information origins, learning does not lead to a collectively intelligent behaviour. This effect is reversed when at least some of the information has as stable origin. Additionally, when the turbulence of the environment is very high learning produces a collectively intelligent behaviour only for high levels of inter-group connectivity. 


\subsection{Implications for Information Management}
Understanding the learning mechanism and its interplay with environmental uncertainty, environmental turbulence, and the number of inter-group ties available can provide useful insights for the design of information management strategies that foster inter-group information exchange and collective intelligence in volatile environments. Two main indications for policy can be drawn from this study respectively related environmental uncertainty and turbulence.

Concerning environmental uncertainty and the associated stability of information origin, this research shows that learning increases the number of IBSs that emerge and the external information they retrieve only in the case of stable origin. However, in this chaotic and unpredictable world marked by increasing volatility, the environments in which groups operate are often characterized by high levels of environmental uncertainty and thus information with an unstable origin (shocks).  
As such, effective strategies for managing this type of information are required to support effective information exchange across groups that fosters collective intelligence in volatile environments. If information with an unstable origin can be re-directed to stable origins which consistently provide it to other actors, this will enable the actors develop an information collection preference for this source through learning and more effectively retrieve the external information needed. Such a change in stability could be achieved by gathering shocks through crowdsourcing and sharing them widely through a fixed node in the information exchange network such as an online platform, website, or social media account (thus establishing a stable information origin) \citep{holderness2015social}. 

With regards to environmental turbulence, a higher number of inter-group ties was consistently found to be a key element in fostering the emergence of IBSs, especially at high levels of environmental turbulence and at any level of environmental uncertainty. To enable the availability of such inter-group ties it is key to build trusted relationships across groups that can be leveraged when external information needs to be retrieved by the group from its environment, also known as bridging social capital \citep{claridge2018functions}. This aligns with previous research indicating that bridging social capital supports inter-group information exchange and resilience in volatile settings such as disaster response \citep{hawkins2010bonding, aldrich2015social, tasic2016informational} and supply chain operations \citep{pettit2010ensuring, JIA2020101614, golgeci2020does}. Additionally, this study shows the importance of the degree of bridging social capital (measured by the number of trusted inter-group ties available) in facilitating effective information exchange via emergent IBSs, even amidst high volatility and uncertainty. Thus, policy interventions should prioritize establishing bridging social capital through initiatives that build trusted connections among different groups, such as between communities and professional responders \citep{norris_community_2008, agger2015area}.

\subsection{Future research}

This study focused on theory building by advancing propositions concerning the emergence of IBSs that effectively convey information across groups in volatile environments. These propositions are designed based on the literature, empirically grounded in the case study of disaster response in Jakarta, and systematically explored through an empirical ABM \citep{nespeca_methodology_2023}. Despite this grounding, the propositions still result from exploratory research on one case study. As such, generalising the propositions will require further investigation and testing e.g. via additional case studies and experiments. Future studies should extend beyond disasters (as in this research), to include other volatile environments such as supply chains during rapid market shifts.

Further, this study focused on the emergence of boundary spanners, and was agnostic to the specific information technology that was used. At the same time, evidence suggests that information technology can play a key role in information sharing and the emergence of inter-group information exchange in volatile environments \citep{tim2017digitally} and also interplay with the emergence of IBSs \citep{levina2005emergence, vanOsch2016team}. Such an interplay, is still poorly understood in volatile environments and requires further research. 

Additionally, agents in this study exhibit non-strategic behavior in information exchange, lacking consideration for long-term goals or personal agendas (i.e., they are myopic, cf. Appendix A).  However, previous research indicates that strategic information exchange occurs, for example, to persuade others to reciprocate with valuable information, to obfuscate or withhold information, or to spread misinformation to advance personal or organizational interests \citep{heavey2020strategic, comes2020coordination, nespeca2020towards}. This strategic sharing can alter recipients' information collection preferences and impact the flow of information, thereby influencing the emergence of IBSs. Future research should investigate how strategic sharing, combined with learning and adaptation in information exchange preferences, affects the emergence of effective IBSs.

Next, this study assumes a constant information exchange network, however, establishing new connections to retrieve relevant information external to a group is often considered as one of the tasks carried out by boundary spanners \citep{ancona1990beyond, ancona1992bridging, marrone2007multilevel}. As such, the network can also change and develop over time. This is for example the case in the context of international humanitarian response operations in which, due to the high staff turnover, few connections across groups enabling inter-group information exchange are available and new ones need to be established ad-hoc during the humanitarian response \citep{altay2014challenges}. 
\citet{zagorecki_interorganizational_2009} studies the emergence of inter-group information exchange when enabling actors to establish connections that decay over time when unused. Their study finds that these networking activities lead to the emergence of actors that convey information across groups. However, the focus of their research is placed on the quantity of information rather than on its quality (e.g., relevance of the information exchanged), and it does not measure the emergence of effective IBSs at the micro level. 
As such, further research is required to study the impact of networking activities and contact decay on the emergence of effective IBSs in volatile environments from the micro level perspective. 

Finally, the adaptation of information exchange preferences over time can not only lead to the emergence of IBSs that facilitate information exchange across groups but also to the formation of information exchange bubbles that isolate groups from others leading to fragmentation and a lack of collective intelligence \citep{comes2020coordination}. 
In other cases, fragmentation in the form of pre-defined functional division of routine tasks (or differentiation) among different organizations and groups (e.g., police and fire fighters) can be beneficial for operational efficiency and collective intelligence as it enables to operate according to pre-established standards and procedures that require less integration and information exchange \citep{lawrence1967organization}. Such fragmentation can, however, be lost in volatile contexts as actors and their groups adapt their activities to a continually and unpredictably changing environment \citep{wolbers2018introducing, nespeca2020towards}. In such situations, inter-group information exchange via IBSs is necessary to reinstate this fragmentation and support collective intelligence \citep{wolbers2018introducing}. Further research is required to study the interplay between the emergence of IBSs on one hand and the formation of information bubbles, fragmentation, and collective intelligence on the other.

\section{Conclusions}
In this increasingly chaotic and rapidly-changing world different groups including governmental and non-governmental organizations, and communities need to work together effectively while operating in volatile conditions (characterized by high turbulence and uncertainty). To this end, the prompt exchange of vital information concerning environmental change across groups is crucial to support coordination and collective intelligence. This article aimed to propose a method to measure the emergence of Informational Boundary Spanners (IBSs) and their effectiveness in fostering inter-group information exchange. Further, a novel Agent-Based Modeling (ABM) framework was introduced to systematically study mechanisms that lead to the emergence of IBSs. The proposed method and ABM are then used to create new insight into one specific mechanisms that explain the emergence of IBSs in volatile environments: i.e, learning.  

Our results show that learning leads to the emergence of effective IBSs when the information needed has a relatively stable origin (i.e. it is consistently provided by the same node in the network). Further, a highly turbulent and volatile environment can easily lead to informational overload. In this situation, retrieving and sharing all relevant information needed becomes challenging. This paper shows that the availability of several contacts (here 20 or more) that can share information across the groups is essential for facilitating the emergence of more IBSs, which helps distribute the load of inter-group information exchange and improves the effectiveness of such exchange. Moreover, when the inter-group contacts are numerous (20 or more), and the level of environmental turbulence is high (above 10 disruptive events per day), the performance of IBSs in facilitating inter-group information exchange is increased with learning. 

Implications of this study include the possibility to use the proposed method and ABM framework to investigate and understand mechanisms for the emergence of informational boundary spanning through a combination of case study research and agent-based simulation. Further, in promoting the emergence of IBSs through learning, actors from different groups exhibit collectively intelligent behavior by choosing agents that effectively facilitate information exchange across groups. This collective intelligence is contingent on the volatility of the environment and requires stable sources of information and a high density of inter-group ties to be effective.
Finally, policy implications of this research  consist of the need to (a) collect and summarize information from unstable origins (e.g., via crowdsourcing) and release such information from stable sources so that agents can learn where to find the information they need and (b) build and maintain trusted connections among groups to ensure collectively intelligent behaviour and effective inter-group information exchange, and support coordination and collaboration even for high levels of volatility. 

Further research will focus on investigating the interplay between the emergence of boundary spanning, learning, and the use of information technology, strategic information exchange behaviour (e.g. spread of misinformation, obfuscation, and persuasion to share back), and the formation of information exchange bubbles and fragmentation. 

\section{Acknowledgements}
 The authors thank Dr. Vìtor V. Vasconcelos for his helpful suggestions. 

\section{Funding}
This work was partly funded by the COMRADES project Grant agreement No 687847, under the EU’s Horizon 2020 research and innovation programme.

\section{Declaration of conflicting interests}
The author(s) declared no potential conflicts of interest with respect to the research, authorship, and/or publication of this article.

\begin{figure*}[h!]
    \centering
    \includegraphics[width=\textwidth]{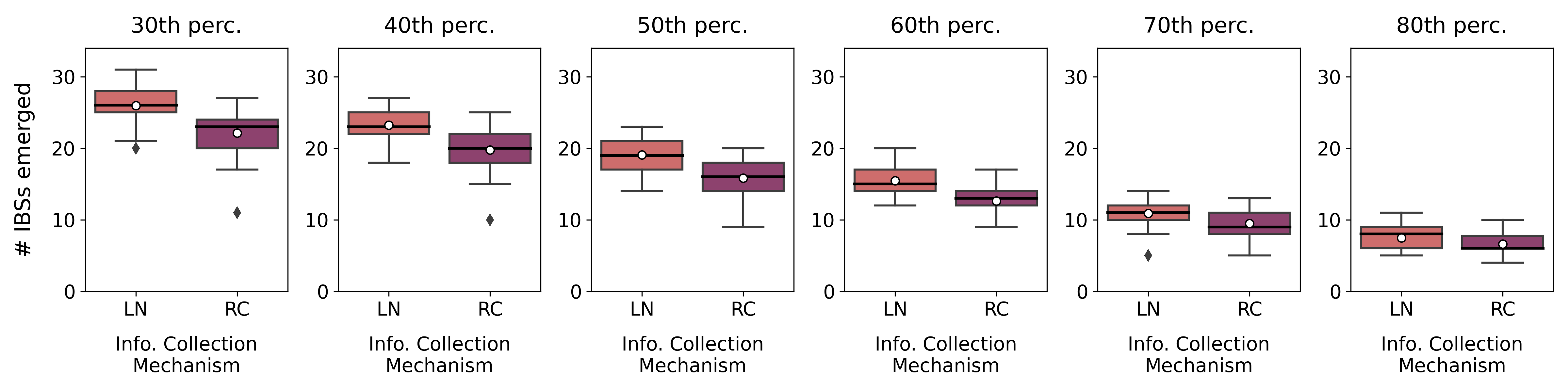}
    \caption{Experiment 1: comparison of the number of emergent IBSs obtained with different thresholds ranging from the 30th (first figure on the left) to the 80th  (last figure on the right) percentiles for the two information collection mechanisms LN (LearNing) and RC (Random Collection).
    }
    \label{fig:exp-1_thresholds}
\end{figure*}

\section*{Appendix A: Learning Mechanism}
To implement the learning mechanism introduced in the section "\textit{Mechanisms for the emergence of informational boundary spanners}" and mentioned in Proposition 2 this study relies on a reinforcement learning algorithm.  Specifically, Q-learning \citep{watkins1989learning, watkins1992q} was chosen given it enables agents to develop information collection preferences over time through a series of information collection activities, as introduced in Proposition 1. The following formula is used to update the expected reward resulting from collecting information from each of the agents' contacts. 

\begin{equation}
   Q(s_t, a_t) \leftarrow (1 - \alpha) \cdot Q(s_t, a_t) + \alpha \cdot R(s_{t + 1}, a)
\end{equation}

Where: 
\begin{itemize}
    \item $Q(s_t, a_t)$ = expected reward for the action a (collecting information from a particular contact) at a given observed state of the environment (responding to a disaster). In practice, Q represents the extent to which the agent expects that a particular contact will provide information that can address its information needs. 
    \item $\alpha$ = learning rate determining the relative importance of the quality of the new information provided by contacts compared to the quality of the information provided so far by the contacts. 
    \item $R(s_{t + 1}, a)$ = actual reward obtained by the agent through carrying out the information collection action $a$ given its information collection state $s$. In this study, this reward is $1$ when the collected information addresses the collecting agent's information needs, and $0$ otherwise. 
\end{itemize}

The agents are assumed to be myopic, meaning that they do not consider the strategic pursuit of long-term high rewards. Rather, the agents simply consider current rewards in their learning process. As such, the value of the discount factor is equal to 0 (and thus not displayed in the equation above).

An agent's \textit{information collection preferences} are represented by the probabilities of the agent choosing each of its contacts as its information collection source. Such probabilities are computed for each contact as the ratio of the Q value associated with the contact, divided by the sum of all Q values associated with all of the agent's contacts. As such, a higher Q value (expected reward) compared to other contacts, entails that the agent will be more likely to choose such contact among the others when collecting information. 

\section*{Appendix B: studying emergent IBSs with different thresholds}

Experiment 0 showed that multiple thresholds enable to individuate IBSs candidates that significantly contribute to inter-group information exchange, thus qualifying as emergent IBSs. Further analysis is needed to determine if such different thresholds yield the same results in the study of IBSs emergence. In this case, providing the same results means that the conclusions regarding IBSs emergence remain the same, i.e. the propositions remain supported by experimental results independently from the thresholds adopted. 

In this appendix, the data from Experiments 1 to 3 is analyzed through six different thresholds deemed adequate to capture emergent IBSs, namely the 30th, 40th, 60th, 70th, and 80th percentiles in the FEs distribution (cf. Experiment 0 - Results Section). The following paragraphs compare the results obtained with these thresholds for each experiment. The appendix considers only the number of emergent IBSs and not their effectiveness in fostering inter-group information exchange. This is because IBSs effectiveness is independent from the threshold adopted. 

Figure \ref{fig:exp-1_thresholds} shows the results of Experiment 1 and illustrates a comparison between the influence of LN compared to RC on the emergence of IBSs when adopting different thresholds. Despite quantitative variations, LN consistently produces more IBSs than RC, thereby supporting Proposition 1 regardless of the threshold used.

\begin{figure*}[h!]
    \centering  
    \includegraphics[width=0.7\textwidth]{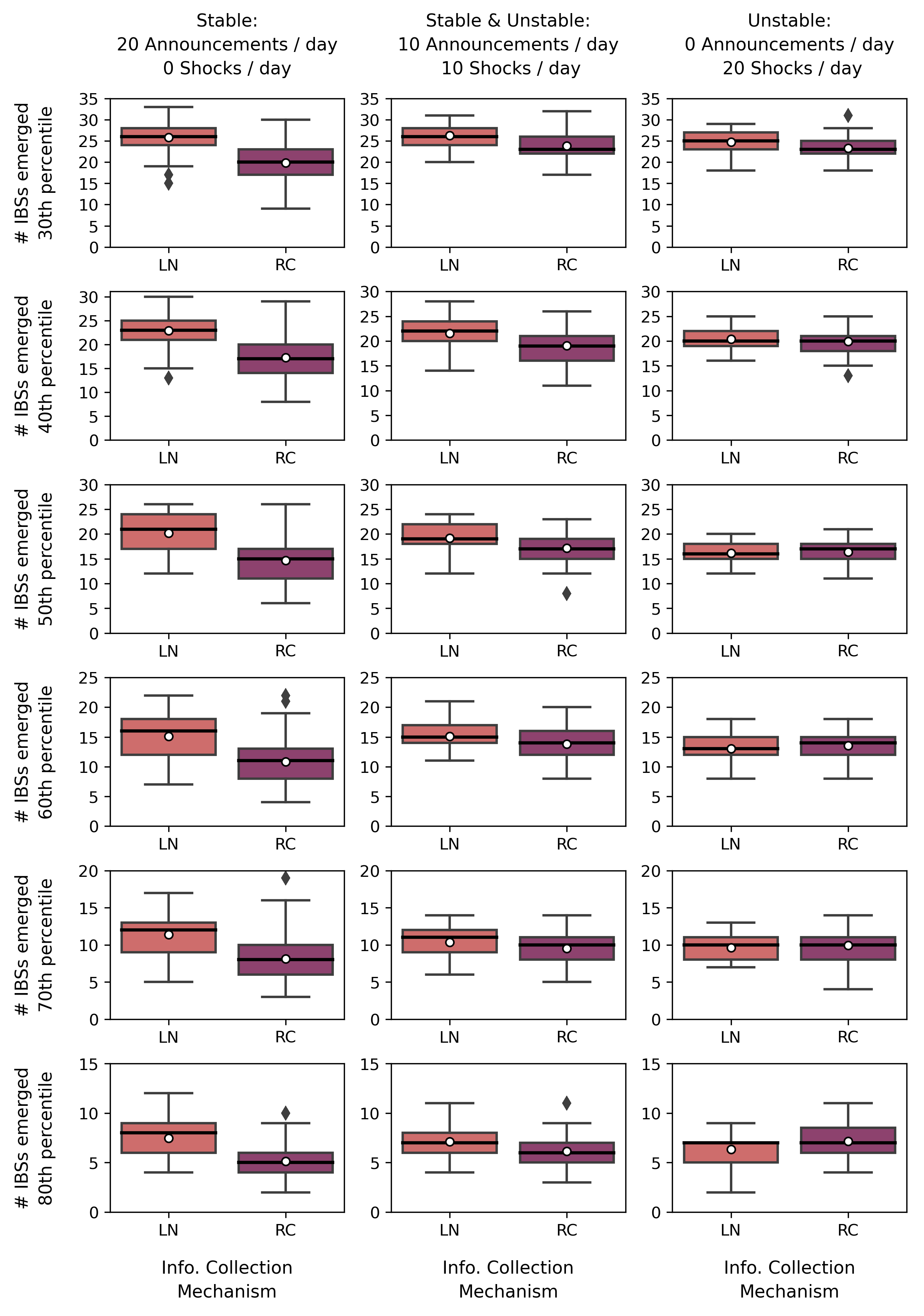}
    \caption{Experiment 2: comparison of the number of emergent IBSs obtained for LearNing (LN) and Random Collection (RC) with the 30th, 40th 50th, 60th, 70th, and 80th percentiles (each represented by one row) and different numbers of stable and unstable information becoming relevant for the groups during each day of simulation (captured in the three columns). While the composition is different, the total number of daily relevant information remains constant at 20 pieces of information per day. }
    \label{fig:exp-2_thresholds}
\end{figure*}

Figure \ref{fig:exp-2_thresholds} presents the results of Experiment 2, comparing the number of emergent IBSs observed with different thresholds for RC and LN when varying the stability of the information origin. In the figure, each row represents the results obtained with one threshold, while the columns represents varying levels of stability, namely stable (only announcements), a combination of unstable and stable (both announcements and shocks), and unstable (only shocks) information. As can be observed by comparing the figures in each row from the from left to right, an increasing number of shocks (unstable information origin) compared to the number of announcements (stable information origin) reduces the impact of learning on fostering the emergence of more IBSs. This observation supports proposition 2 and holds across all thresholds. 

Further, with only unstable information (right column), learning minimally impacts the number of emerged IBSs, with effects varying by threshold. From the 40th to the 70th percentiles, LN has no significant influence on IBS emergence. However, the lowest and highest thresholds (respectively the 30th and 80th percentiles) produce different and conflicting results. Precisely, with the 30th percentile, LN increases IBS emergence, but with the 80th percentile, this effect reverses. In this case, these discordant results are considered outliers and disregarded for two reasons. First, they lead to opposite conclusions, likely because they are the most extreme among those found to be adequate. Second, all of the other thresholds (four out of six) consistently produced similar results, confirming that the discrepancies observed with the 30th and 80th percentiles are due threshold selection rather than experimental data. Consequently, results from the 30th and 80th thresholds are excluded. The remaining findings from the 40th, 50th, 60th, and 70th percentiles consistently indicate that learning has no impact on the emergence of IBSs with unstable information origins, supporting Proposition 2.  

\begin{figure*}[h!]
    \centering \includegraphics[width=0.8\textwidth]{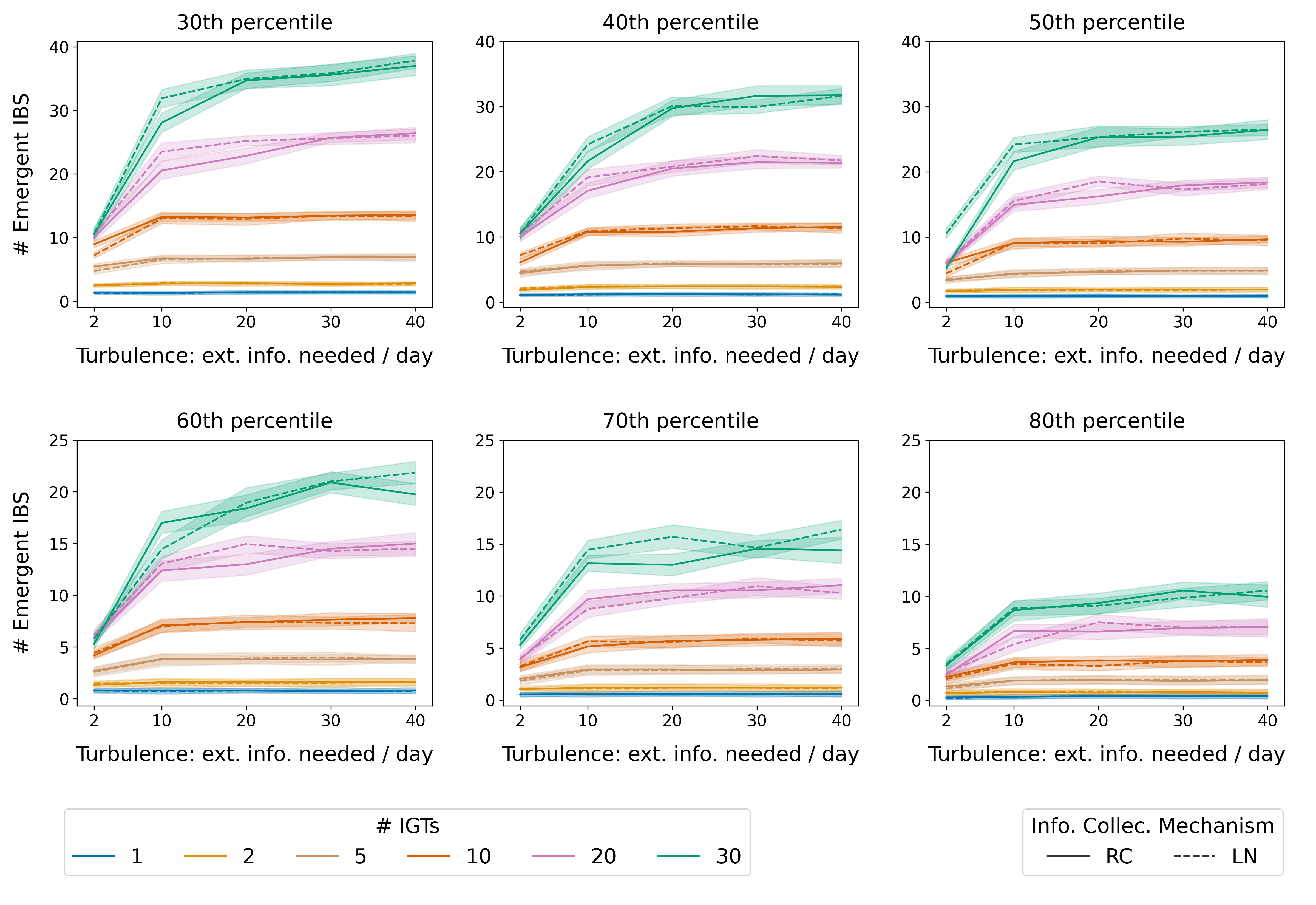}
    \caption{Experiment 3: comparison of the number of emergent IBSs obtained with LearNing (LN) and Random Collection (RC) for the 30th, 40th, 50th, 60th, 70th, and 80th percentiles,  different levels of Turbulence, and numbers of Inter-Group Ties (\# IGTs).}
    \label{fig:exp-3_thresholds_new}
\end{figure*}

Finally, Figure \ref{fig:exp-3_thresholds_new} presents the results of experiment 3 and shows a comparison in the number of emergent IBSs with LN and RC for varying levels of environmental turbulence and inter-group ties. This figure illustrates that for all thresholds the number of IBSs emerged grows with the number of inter-group ties and the level of environmental turbulence. Further, the number of IBSs emerged is not significantly affected by the information collection mechanism adopted. These findings support proposition 3 and are evident independently from the threshold adopted. 

This appendix demonstrates that using various adequate thresholds, as established in Experiment 0, leads to consistent conclusions across Experiments 1 to 3, supporting Propositions 1, 2, and 3 regardless of the threshold used. It highlights the importance of comparing and assessing the consistency of findings across different thresholds when using the method to study the emergence of IBSs. This is necessary as, in specific instances like the 30th and 80th percentiles in Experiment 2, different thresholds may yield conflicting results. Such discrepancies can lead to reconsider and revise conclusions regarding the emergence of IBSs. In this case, the discordant results with the 30th and 80th were considered outliers and disregarded given their extreme values and inconsistent results with the majority of the other thresholds. Therefore, the conclusions of Experiment 2 remained unchanged.

\bibliographystyle{SageH}
\bibliography{bibliography.bib}

\end{document}